\documentclass[11pt,letterpaper]{article}
\usepackage[T1]{fontenc}
\usepackage{babel}
\usepackage{verbatim}
\usepackage{bm}
\usepackage{amsmath,amstext,amsthm,amssymb}
\usepackage{float}
\usepackage{xspace}
\usepackage{graphicx}
\usepackage{mathtools}
\usepackage{subcaption,array}
\usepackage{wrapfig}
\usepackage[noend]{algorithmic}
\usepackage{booktabs}
\usepackage{makecell}
\usepackage[dvipsnames]{xcolor}
\definecolor{DarkGray}{rgb}{0.66, 0.66, 0.66}
\definecolor{DarkPowderBlue}{rgb}{0.0, 0.2, 0.6}
\definecolor{fluorescentyellow}{rgb}{0.8, 1.0, 0.0}

\usepackage{tikz-qtree,tikz-qtree-compat}
\usepackage{tikz}

\usepackage[unicode=true,
bookmarks=false,
breaklinks=false,pdfborder={0 0 1},backref=none,colorlinks=true,allcolors=blue]
{hyperref}

\usepackage{cleveref}
\usepackage{thmtools}
\usepackage{thm-restate}
\usepackage{bbm}
\usepackage[bottom]{footmisc}
\usepackage[font={small,it}]{caption}

\usepackage[ruled,vlined,linesnumbered,algonl]{algorithm2e}
\SetEndCharOfAlgoLine{}
\SetKwComment{Comment}{\footnotesize$\triangleright$\ }{}

\SetCommentSty{mycommfont}

\Crefname{algocf}{Algorithm}{Algorithms}
\crefname{algocfline}{line}{lines}
\Crefname{invariant}{Invariant}{Invariants}
\Crefname{claim}{Claim}{Claims}
\Crefname{subclaim}{Subclaim}{Subclaims}

\usepackage{enumitem}
\usepackage{fullpage}
\usepackage{nicefrac}
\usepackage[compact]{titlesec}

\usepackage{setspace}

\makeatletter
\allowdisplaybreaks
\makeatother

\declaretheorem[name=Theorem,numberwithin=section]{theorem}
\declaretheorem[sibling=theorem]{lemma}
\declaretheorem[sibling=theorem]{claim}

\declaretheorem[name=Corollary,sibling=theorem]{cor}

\theoremstyle{definition}
\declaretheorem[name=Definition,sibling=theorem]{defn}
\declaretheorem[sibling=theorem]{remark}

\declaretheorem[name=Observation,sibling=theorem]{obs}

\DeclareMathOperator*{\argmin}{arg\,min}

\ifx\proof\undefined
\newenvironment{proof}[1][\protect\proofname]{\par
	\normalfont\topsep6\p@\@plus6\p@\relax
	\trivlist
	\itemindent\parindent
	\item[\hskip\labelsep\scshape #1]\ignorespaces
}{%
	\endtrivlist\@endpefalse
}
\providecommand{\proofname}{Proof}
\fi

\usepackage{babel}

\usepackage[textsize=tiny,textwidth=2cm,color=green!50!gray,obeyFinal]{todonotes} 

\newcommand{\alert}[1]{{\color{red}#1}\marginpar{$\star\star$}}

\newcommand{\anupam}[1]{\textcolor{red}{AG: #1}\marginpar{$\star\star$}}

\newcommand{\madhu}[1]{\textcolor{blue}{Madhu: #1}\marginpar{$\star\star$}}
\newcommand{\guru}[1]{\textcolor{ForestGreen}{Guru: #1}\marginpar{$\star\star$}}

\usepackage{soul}
\sethlcolor{fluorescentyellow}

\makeatother

\newcommand{\norm}[1]{\left\lVert#1\right\rVert}
\newcommand{\sumL}{\sum\limits}
\newcommand{\nf}{\nicefrac}

\newcommand{\BB}{\mathbb{B}}

\newcommand{\EE}{\mathbb{E}}

\newcommand{\RR}{\mathbb{R}}

\newcommand{\ZZ}{\mathbb{Z}}

\newcommand{\cA}{\mathcal{A}}
\newcommand{\cB}{\mathcal{B}}
\newcommand{\cC}{\mathcal{C}}

\newcommand{\cE}{\mathcal{E}}
\newcommand{\cF}{\mathcal{F}}

\newcommand{\cH}{\mathcal{H}}
\newcommand{\cI}{\mathcal{I}}

\newcommand{\cM}{\mathcal{M}}

\newcommand{\cP}{\mathcal{P}}

\newcommand{\cV}{\mathcal{V}}
\newcommand{\cW}{\mathcal{W}}
\newcommand{\cX}{\mathcal{X}}
\newcommand{\cY}{\mathcal{Y}}

\newcommand{\vol}{\operatorname{Vol}}

\newcommand{\dist}{\operatorname{dist}}

\newcommand\cone{\operatorname{cone}}
\newcommand\envelope{\operatorname{env}}

\newcommand{\ip}[1]{\langle #1 \rangle}

\newcommand{\eps}{\varepsilon}
\newcommand{\sse}{\subseteq}

\newcommand{\poly}{\operatorname{poly}}
\newcommand{\elts}{U}

\newcommand{\CDfull}{conic dimension\xspace}
\newcommand{\CDshort}{conic dimension\xspace}
\newcommand{\CDmath}{\operatorname{ConicDim}}

\newcommand{\eat}[1]{}

\newcommand{\codim}{s}

\newcommand{\tO}{\widetilde{O}}

\usepackage{tcolorbox}
\tcbuselibrary{theorems}
\newtcbtheorem[number within=section]{myquest}{Question}%
{colback=red!5,colframe=red!35!black,fonttitle=\bfseries}{th}

\usepackage[margin=1in]{geometry} 
\usepackage{enumitem} 

\title{\vspace{-1.5cm} Combinatorial Optimization using Comparison Oracles}

\author{
Vincent Cohen-Addad$^*$ \and
Tommaso d'Orsi$^\dagger$ \and
Anupam Gupta$^\ddagger$ \and
Guru Guruganesh$^*$ \and
Euiwoong Lee$^\S$ \and
Renato Paes Leme$^*$ \and
Debmalya Panigrahi$^\P$ \and
Madhusudhan Reddy Pittu$^\ddagger$ \and
Jon Schneider$^*$ \and
David P. Woodruff$^\|$
}

\date{\vspace{-1.5cm}}

\begin{document}
\thispagestyle{empty}
\maketitle

\begingroup
\renewcommand\thefootnote{}\footnote{
$^*$Google Research. Emails: \texttt{\{vcohenad,gurug,renatoppl,jschnei\}@google.com} \quad
$^\dagger$Bocconi University. Email: \texttt{tommasodorsi@gmail.com} \quad
$^\ddagger$New York University. Emails: \texttt{\{anupam.g,madhusudhan.p\}@nyu.edu} \\
$^\S$University of Michigan. Email: \texttt{euiwoong@umich.edu} \quad
$^\P$Duke University. Email: \texttt{debmalya@cs.duke.edu} \quad
$^\|$Carnegie Mellon University. Email: \texttt{dwoodruf@andrew.cmu.edu}
}%
\addtocounter{footnote}{-1}%
\endgroup

\begin{abstract}
  In a linear combinatorial optimization problem, we are given a family $\mathcal{F} \subseteq 2^U$ of \emph{feasible subsets} of a ground set $U$ of $n$ elements, and our goal is to find $S^* = \arg\min_{S \in \mathcal{F}} \langle w,\mathbbm{1}_S \rangle$. Traditionally, we are either given the weight vector up-front, or else we are given a \emph{value oracle} which allows us to evaluate $w(S) := \langle w, \mathbbm{1}_S \rangle$ for any $S \in \mathcal{F}$. Motivated by recent practical interest in solving problems via pairwise comparisons, and also by the intrinsic theoretical quest to understand fundamental computational models, we consider the weaker and more robust \emph{comparison oracle}, which for any two feasible sets $S, T \in \mathcal{F}$, reveals only if $w(S)$ is less than/equal to/greater than $w(T)$. We ask:
  \begin{quote}
    When can we find the optimal feasible set $S^* = \arg\min_{S \in \mathcal{F}} w(S)$ using a small number of comparison queries? If so, when can we do this efficiently?
  \end{quote}
  We present three main contributions:
  \begin{enumerate}[noitemsep, topsep=2pt, parsep=2pt]
    \item Our first main result is a surprisingly general answer to the query complexity. We establish that the query complexity for the above problem over any arbitrary set system $\mathcal{F} \subseteq 2^U$ is $\tO(n^2)$. This result leverages the inference dimension framework, and demonstrates a fundamental separation between information complexity and computational complexity, as the runtime may still be exponential for NP-hard problems (assuming the Exponential Time Hypothesis).

    \item Having resolved the query complexity, we turn to the algorithmic question. We develop two general algorithmic frameworks:
    \begin{enumerate}[noitemsep, topsep=2pt, parsep=2pt]
        \item \textbf{Optimization from Certification:} We present a novel Dual Ellipsoid framework that establishes an efficient reduction from optimization to certification. This framework demonstrates that to optimize efficiently, it is sufficient to design an efficient certification for the optimality of a candidate set $S^*$ with the knowledge of $w^*$ using only comparisons between feasible sets. Moreover, this framework also yields a deterministic low query complexity algorithm. 
        \item \textbf{Global Subspace Learning (GSL):} Tailored for integer objective functions bounded by $B$, we sort all feasible sets using only $O(nB \log(nB))$ queries, improving upon the $\tO(n^2)$ bound when $B=o(n)$. We efficiently implement this framework for linear matroids via algebraic techniques, yielding efficient algorithms with improved query complexity for problems like $k$-SUM, SUBSET-SUM, and $A+B$ sorting.
    \end{enumerate}

    \item Our third set of results gives the first polynomial-time, low-query algorithms for several classic combinatorial problems. We develop such algorithms for finding minimum cuts in simple graphs, minimum weight spanning trees (and matroid bases in general), bipartite matching (and matroid intersection), and shortest $s$-$t$ paths. 
\end{enumerate}
  Our work provides the first general query complexity bounds and efficient algorithmic results for this fundamental model.  
\end{abstract}
\newpage
\setcounter{page}{1}

\section{Introduction}
\label{sec:introduction}

In a (linear) combinatorial optimization problem, we are given a
family $\mathcal{F} \subseteq 2^U$ of \emph{feasible subsets} of a
ground set $U$ of $n$ elements, and our goal is to find
$S^* = \arg\min_{S \in \mathcal{F}}
\ip{w,\mathbbm{1}_S}$. Traditionally, we are either given the weight
vector $w$ explicitly, or else we are given a \emph{value oracle}
which allows us to evaluate $w(S) := \ip{w, \mathbbm{1}_S}$ for any
$S \in \cF$.  In this work, we pose the question:
\begin{quote}
  Which combinatorial optimization problems can we efficiently solve
  if we are only allowed \emph{pairwise comparisons between feasible
    solutions?}
\end{quote}
That is, we are given access to a comparison oracle that takes two
\underline{feasible} sets $S, T \in \cF$, and outputs whether $w(S)$
is less than/equal to/greater than $w(T)$. This comparison query model
is perhaps the weakest feedback model where we can hope to optimize
efficiently.\footnote{One can consider more general objective
  functions, such as XOS, submodular, subadditive, etc., where the
  function $w(\cdot)$ is typically exponential-sized, and hence we can
  only access it via value or demand oracles. Our question---what can
  you do with just comparison queries---is interesting in these
  settings as well, but we focus on linear optimization for now.}
As a concrete question, consider the min-cut problem in this model. Specifically:
\begin{quote}
  We are given an undirected simple graph $G = (V,E)$. In each query, we can compare any two (edge) cuts to each other. \emph{Can we efficiently find a minimum cut in this graph?}
\end{quote}
The min-cut problem has been intensively studied in the value-query
model in recent years, leading to new insights and algorithms: can we
replicate this success in the comparison-query model?

More generally, our high-level motivation for the model comes from recent interest in applications (e.g., from crowdsourcing, recommendation systems, and most recently, RLHF) which solicit feedback from users only via comparison queries. In these
settings, users' notions of quality/utility is often difficult to
quantify, or noisy, or only partially consistent, and hence they are
asked to give \emph{relative preferences}---choosing one alternative
over another---rather than assigning precise numerical values. While
these practical settings do not seek to solve precise optimization
problems (like min-cuts or spanning trees), they do motivate the
particularly clean query model we investigate in this work.

\subsection{Our Results}
\label{sec:our-results}

\subsubsection{Query Complexity}
\label{sec:intro-query-complexity}

We first focus on the \emph{query complexity} of the problem: how many
comparison queries give us enough information to solve the problem,
without worrying about computation? Recall that we can only query
pairs of feasible solutions, so it is completely unclear whether we
can identify the optimal solution using much less than $|\cF|$
queries.

For this query complexity question, we show a surprisingly general
\emph{positive} result: \emph{we can optimize over \underline{all
    Boolean families} with a small number of queries!}
\begin{restatable}[Boolean Linear Optimization]{theorem}{BLO}
	\label{thm:query-comp-boolean}
	For any family $\mathcal{F} \sse 2^U$ and unknown weight
        function $w^*: U \rightarrow \RR$, we can solve
        $\arg\min_{S\in \cF} \sum_{e \in S} w^*_e$ using
        $O(n \log^2 n \cdot \log |\mathcal{F}|) = \widetilde{O}(n^2)$
        comparison queries, where $n = |U|$.
\end{restatable}
We find \Cref{thm:query-comp-boolean} remarkable: it shows that the
number of comparison queries required to find the optimal solution is
only $\widetilde{O}(n^2)$, \emph{regardless of the complexity of} the
set system $\mathcal{F}$. Indeed, we could consider $\mathcal{F}$ to
represent set covers, independent sets, cliques, or other NP-hard
problems for which finding the optimal set is believed to be
computationally hard---nonetheless, the number of comparisons required
is still nearly quadratic! Indeed, while the query complexity of the
procedure in~\Cref{thm:query-comp-boolean} is polynomial, its running
time could be exponential (under standard complexity-theoretic
assumptions).  This shows a large gap between the information
complexity and the computational complexity in this model.

\begin{table}[t]
    \centering
    \label{tab:gen-results}
    \renewcommand{\arraystretch}{1.2}
    \begin{tabular}{lcc}
        \toprule
        \textbf{Problem} & 
        \textbf{Past Work} & 
        \textbf{Our Results} \\
        \midrule
        Any Boolean\ & - & $O(n^2 \log^2 n)$ \\
        ~~~~~~optimization problem  & - & $O(nB \log (nB))$ \\
        \bottomrule
    \end{tabular}
    \caption{Our general results for query complexity. The second row is when weights are integers in $[-B,B]$.}
\end{table}

The key to proving \Cref{thm:query-comp-boolean} is to approach 
this query-complexity question from the correct perspective: 
we adapt the powerful
active classification framework of
\cite{KLMZ17-comp-based-classification} to the optimization
setting. We show how to use their classification framework to also optimize over
arbitrary point sets (and not just Boolean point sets) in
$\mathbb{R}^d$, with the number of queries being related to their
``\CDfull'':\footnote{The \CDfull is a slightly weaker variant of the ``inference dimension'' from \cite{KLMZ17-comp-based-classification}, specialized for linear optimization. The name reflects its reliance on the notion of conic independence (based on conic spans).}
\begin{restatable}[General Linear Optimization]{theorem}{GLO}
  \label{thm:query-complexity-conic-dimension}
  There is an algorithm that, for any point set $\cP \sse \RR^d$ with
  \CDfull $k$ and unknown weights $w^* \in \RR^d$, returns the minimizer
  $x^* = \arg\min_{x \in \cP} \ip{w^*, x}$ using at most
  $O(k\log k\log |\cP|)$ comparisons.
\end{restatable}
Since the \CDfull of the Boolean hypercube $\{0,1\}^n$ is known to be
$O(n\log n)$, this result immediately
implies~\Cref{thm:query-comp-boolean}. Moreover, it gives bounds on
the query complexity when the point sets have bounded representation
size, and for approximate optimization when they have bounded norm. We
discuss these results in~\Cref{sec:gen_set_system}.

\subsubsection{Optimization from Certification}

While the approach of \Cref{thm:query-comp-boolean} provides strong query complexity bounds, we also provide a powerful alternative perspective via certification. In \Cref{sec:ellipsoid}, we show an efficient reduction from optimization to certification using a Dual Ellipsoid Algorithm.  This framework reduces the optimization problem to designing a Conic-Certification Oracle (CCO). If one can efficiently certify the optimality of a candidate solution (with the knowledge of $w^*$) using only comparisons, one can efficiently optimize. Moreover, this framework also yields a deterministic algorithm that uses $\poly(n)$ queries for the Boolean Linear Optimization problem. 

\begin{theorem}[Optimization from Certification]\label{thm:dual-ellipsoid-intro}
There is a deterministic algorithm that, given access to a Conic-Certification Oracle $\mathrm{(CCO)}$ for a family $\mathcal{F}$ represented by points $\mathcal{P} \subseteq \mathbb{R}^d$ with bit complexity $\langle \mathcal{P} \rangle$, finds the optimal solution $S^* = \arg\min_{S \in \mathcal{F}} \langle w^*, \mathbbm{1}_S \rangle$ using $O(d^3 \langle \mathcal{P} \rangle)$ calls to the $\mathrm{CCO}$ and separating hyperplane updates.
\end{theorem}

See \Cref{sec:query-compl-runn} for the proof of \Cref{thm:dual-ellipsoid-intro}. The fact that this framework also yields a deterministic algorithm that uses $\poly(n)$ queries for the BLO problem follows from \Cref{cor:DE-hypercube-qc}.
We demonstrate the power of this reduction by using it to efficiently find minimum weight perfect matchings in complete bipartite graphs.

\subsubsection{Global Subspace Learning}
\label{sec:intro-glob-subsp-learn}
Alongside the certification reduction, we develop a second, general algorithmic tool tailored for the case where the weights in the objective function are integers in the range $[-B, B]$. This Global Subspace Learning (GSL) algorithm is highly intuitive, and in fact, sorts all the feasible sets by their weight (in $\poly(|\cF|, n)$ time for now):
\begin{theorem}[Boolean Linear Optimization: Bounded-Weights]
  \label{thm:bounded-intro}
  There is an algorithm that, for any family $\mathcal{F} \sse 2^U$
  and unknown integer weight vector satisfying $|w^*_e| \leq B$, sorts
  the feasible sets according to their weight $w^*(S)$ (and hence solves
  the optimization problem) using $O(nB \log (nB))$ comparison
  queries, where $n = |U|$. 
\end{theorem}

The main insight behind this result is the following: two sets $S, T\in \cF$ with the
same weight satisfy 
$\langle w^*, \mathbbm{1}_S-\mathbbm{1}_T\rangle =0$, and hence the
vector $(\mathbbm{1}_S-\mathbbm{1}_T)$ gives us information about the
subspace orthogonal to $w^*$. Our algorithm maintains a subspace $\cA$
spanned by these orthogonal directions
$(\mathbbm{1}_S-\mathbbm{1}_T)$, and uses it to infer the weight class
of other sets in $\cF$ without making any additional comparisons;
the details appear in \Cref{sec:GSL}. While \Cref{thm:bounded-intro} relies
on weights being integral and bounded, it highlights how the structure
of linear optimization can be exploited.


This explicit linear-algebraic structure allows us to get efficient
algorithms from \Cref{thm:bounded-intro}: 

\begin{restatable}[Efficient Optimization over Linear Matroids]{theorem}{LinMatOpt}
  \label{thm:LinMatOpt}
Let $\cF$ be the bases of a rank-$k$ linear matroid $\cM=(U, \cI)$, represented by $V\in \RR^{k\times n}$ with entry representation size $\poly(n)$. For an unknown integer weight function $w^*$ satisfying $|w^*_e| \leq B$, we can sort the bases according to their weight $w^*(S)$, and hence find the minimum-weight basis $\arg\min_{S\in \cF} w^*(S)$ in $\poly(n, B)$ time. The number of comparison queries required is $O(nB\log nB)$ and more generally, $O((n+C)\log C)$, where $C = |\{w^*(S) : S \in \cF\}|$ is the number of distinct basis weights.
\end{restatable}

\begin{table}[t]
    \centering
    \label{tab:applications}
    \renewcommand{\arraystretch}{1.2}
    \begin{tabular}{lccc}
        \toprule
        \textbf{Problem} & 
        \textbf{Past Work} & 
        \multicolumn{2}{c}{\textbf{Our Results}} \\
        \cmidrule(lr){3-4}
         & & \textbf{Comparisons} & \textbf{Equality queries} \\
        \midrule
        $k$-SUM & $O(kn \log^2 n)$ & $O((n + kB)\log(kB))$ & $O(kB(n + kB))$ \\
        SUBSET-SUM & $O(n^2 \log n)$ & $O(nB \log(nB))$ & $O(n^2 B^2)$ \\
        Sorting $P+Q$ & $O(n \log^2 n)$ & $O((n + B)\log B)$ & $O(B(n + B))$ \\
        Linear Matroids & -  & $O(nB \log(nB))$ & - \\
        \hline
        Min-cuts & - & $\tilde{O}(|V|)$ & - \\
        Graph reconstruction & - & $O(|V|^2, (|E|+|V|) \log |V|)$ & - \\
        Matroids & - & $\tilde{O}(n)$ & - \\
        Matroid Intersection & - & $O(n^4)$ & - \\
        \bottomrule
    \end{tabular}
    \caption{Efficient algorithms and query complexities for problems in the comparison decision tree model. The first set of results are for integer values in $[-B, B]$.
    Our algorithms are efficient, running in $\poly(n,B)$ and $\poly(n)$ time.
    The past work is from \cite{kane2019near}. }
\end{table}


Moreover, we get new algorithms for the \textsc{Subset Sum}, \textsc{$k$-Sum} and \textsc{Sorting $P+Q$} problems studied by \cite{kane2019near}, with better bounds for the case when $B = o(n)$; see \Cref{sec:other-apps} for details.

\subsubsection{Efficient algorithms for Classical Combinatorial Families}

While our general results provide broad query bounds, achieving
computational efficiency for classic combinatorial problems often
requires tailored approaches. We now turn our focus to several such
problems, and show that 
despite the diversity in these optimization tasks, they admit
very efficient algorithms using only comparison queries.

\medskip
\textbf{Graph Cuts.} For the motivating problem of finding min-cuts in
(unweighted) simple graphs, we extend the work of
\cite{RSW18-value-oracle-mincuts} from the more informative
\emph{value oracle} model and match it in the \emph{comparison oracle}
model: 
\begin{restatable}[Minimum cut]{theorem}{mincut}
  \label{thm:unweighted-cuts}
  There is a randomized algorithm that computes the exact minimum cut
  of a simple graph $G$ with high probability using
  $\widetilde{O}(|V|)$ cut comparison queries and
  $\widetilde{O}(|V|^2)$ time.
\end{restatable}
In addition to finding the minimum cut, we show how to recover the
entire edge set of $G$ \emph{deterministic\-ally} using cut comparison
queries, except for some degenerate
cases, 
when the task is provably impossible.
\begin{restatable}[Graph recovery]{theorem}{graphrecov}
  \label{thm:graph_recovery}
  A simple unweighted graph $G\notin \{K_2,\bar{K}_2, K_3,\bar{K}_3\}$
  can be recovered using $O( \min\{(|E|+|V|)\log |V|, |V|^2 \} )$ cut
  comparison queries and in $\widetilde{O}(|V|^2)$ time.
\end{restatable}

Finally, we show that cut comparisons suffice to construct sparsifiers
for a broad class of graph problems called \emph{heavy subgraph}
problems (which include \emph{densest subgraph} and \emph{max-cut}) by
combining our edge-sampling techniques with the results of
\cite{EHW15-uniformsampling}; see \Cref{sec:sparsifier-gen} for
details. 

The min-cut problem illustrates a \emph{qualitative difference between
  the value oracle and comparison oracle settings}: in the former, we
can evaluate the expression $\nf12 (w(\partial u) + w(\partial v) - w(\partial\{u,v\}))$
to learn whether edge $(u,v)$ exists (and what its weight is, if the
graphs are weighted).  In contrast, reconstructing a graph is not
straightforward with cut comparisons. (E.g., all non-trivial cuts in
$K_3$ have the same value, and the same holds for its complement graph
$\bar{K}_3$, so we cannot distinguish them by comparison queries
alone. Moreover, \emph{weighted} graphs cannot be recovered at all using
comparisons, even up to scaling (see \Cref{sec:non-reconstr-graphs})!
Given this, \Cref{thm:graph_recovery} comes as a surprise.

The main idea behind both results is the following: for a vertex $v$
and set $S \not\ni v$, comparing $w(\partial(S+v))$ and $w(\partial(S))$ tells us whether
more than half of $v$'s incident edges go into $S$. We build on this
observation to identify all edges incident to $v$: for a sequence of
nested sets
$\emptyset \neq S_0\sse S_1\sse \ldots \sse S_{n-1} \neq V\setminus
\{v\}$, we ask the above question for each $S_i$ to find a ``tipping
point''---a query where the sign of $w(\partial(S+v)) - w(\partial(S))$
changes---whereupon we find an edge. We then refine the process to
make edge recovery more efficient, and to perform efficient sampling
on $G$ and get a randomized algorithm using only $\tO(n)$ comparison
queries; see \Cref{sec:minimum-cut}.

\medskip\textbf{Weighted Min-Cut.}
These techniques do not extend to weighted graphs. E.g., there are many weighted graphs which cannot be identified using just comparison queries. (E.g., adding tiny-weight edges to a graph changes the graph but not the comparisons.) Also, while our techniques can give \emph{non-adaptive} algorithms using $\poly(n)$ comparisons for min-cut/graph recovery in simple graphs, any non-adaptive algorithm \emph{for the weighted case} must have exponential query complexity in the worst case. (See \Cref{sec:useful-examples} for a discussion of this and other bottlenecks.) 
That said, we give the following result for the case where the graph has a small number of distinct vertex degrees. (See \Cref{sec:weighted-cuts} for a proof.)

\begin{restatable}[Weighted minimum cut]{theorem}{w-mincut}
\label{thm:weighted-cuts}
For a graph $G=(V,E)$ with unknown integer edge weights in the range $[0, B]$, the minimum cut can be found using $(nB)^{O(r^2\log B)}$ queries and time, where $r=|\partial_w(u): u\in V|$ is the number of distinct weighted degrees among all vertices. In the special case of degree-regular graphs, the complexity is $(nB)^{O(\log B)}$.
\end{restatable}

\medskip
\textbf{Matroid Bases, Matroid Intersections, and Paths.}  Next, we
consider other basic combinatorial objects in graphs: \emph{spanning
  trees}, (bipartite) \emph{matchings}, and $s$-$t$ paths. Going
beyond graphs, we also consider natural extensions of the first two
problems to {\em matroid bases} and {\em matroid intersections}.

\begin{theorem}[Matroids, Matchings, and Paths]
  The following combinatorial problems can be solved efficiently in
  the comparison oracle model:
  \begin{enumerate}[label=(\roman*)]
  \item \label{item:mmp-1} the minimum-weight basis of a matroid on $n$ elements can be
    found $O(n \log n)$ comparison queries in $\tO(n^2)$ time,
  \item \label{item:mmp-2} the minimum-weight set in the intersection of two matroids on
    an $n$-element ground set can be found using $O(n^4)$ comparison
    queries/time, and
  \item \label{item:mmp-3} the minimum-length $s$-$t$ path in a graph (or a negative
    cycle, if one exists) can be found using using $O(n^3)$ walk
    comparisons in $O(n^3)$ time.
  \end{enumerate}
\end{theorem}

Note that knowing the order of element weights is enough to optimize
over matroids (using the greedy algorithm), but how can we compare
element weights? We observe that if two elements share a circuit, then
we can find two bases which differ on exactly these elements, and
hence compare them. We then show that one can decompose any matroid so
that we only need to compare such pairs of elements. (See \Cref{sec:matroid-bases}.)

For matroid intersection (and its special case, bipartite matchings),
sorting the element weights is neither doable nor sufficient. Instead,
we implement the shortest augmenting path algorithm (and its natural
extension for matroid intersection) using only comparisons between
matchings (i.e., common independent sets);
see~\Cref{sec:matroid-intersection}. Finally, we show that the
Bellman-Ford algorithm for shortest $s$-$t$ paths can be implemented
using comparisons between $s$-$t$ walks to prove~\ref{item:mmp-3}.

Furthermore, by utilizing the Dual Ellipsoid reduction developed in \Cref{sec:ellipsoid}, we obtain an efficient algorithm for finding optimal perfect matchings in complete bipartite graphs.

\begin{theorem}[Bipartite Perfect Matching]\label{thm:bipartite-pm-intro}
There is a polynomial-time algorithm to find the minimum-weight perfect matching in a complete bipartite graph $K_{n,n}$ using only comparisons between perfect matchings.
\end{theorem}

\paragraph{Organization.} 
In \Cref{sec:gen_set_system}, we present our first result on query complexity for Boolean Linear Optimization. Next, \Cref{sec:ellipsoid} presents our reduction from optimization to certification via the Dual Ellipsoid method. \Cref{sec:GSL} introduces our Global Subspace Learning framework, and gives efficient algorithms for specific classes of set systems. \Cref{sec:minimum-cut} is dedicated to efficient algorithms for finding minimum cuts using cut comparisons. To improve flow, many technical proofs and discussions are deferred to the appendix, which includes the omitted proofs from the Dual Ellipsoid framework (see \Cref{sec: ellipsoid-omitted-proofs}), as well as our algorithms for minimum-weight matroid bases, matroid intersections, and shortest $s$-$t$ paths (see \Cref{sec:comb-algorithms}).

\subsection{Related Work}
\label{sec:related-work}

While our theoretical study of comparison oracles for linear discrete optimization problems is new, to the best of our knowledge, comparison oracles have been studied for
submodular functions and beyond: \cite{balcan2016learning} show how to
use few exact comparison queries to any submodular function to
construct an approximate comparison
oracle. \cite{haghiri2017comparison} studied a comparison model for
metric spaces, where an oracle compares distances of nodes to a fixed
location $i$, and gave small-query algorithms under some expansion
conditions; later work studied a similar access model for
classifica\-tion/regression in Euclidean space
\cite{haghiri2018comparison}.  \cite{xu2017noise} studied binary
classification under noisy labeling and access to a pairwise
comparison oracle. \cite{KLMZ17-comp-based-classification} studied
active classification using label and pairwise comparison queries
between the data points to recover a (linear) classification
boundary. This last result plays an central role in our work on
general linear optimization.  The techniques of \cite{kane2019near}
can be used to derandomize the query strategy for general linear
optimization.

\textbf{Minimum Cuts:}
The study of value (and the more powerful demand) queries has long
been used for general submodular functions, given their natural
representation is exponential sized. The graph min-cut problem can be viewed both as submodular optimization on vertex sets, or linear optimization over the feasible edge sets. Min-cuts have been intensively studied in the value query model in recent years, first by
\cite{RSW18-value-oracle-mincuts} who gave an algorithm using $\tO(n)$
queries in simple graphs. This was extended by
\cite{MN-20-weighted-mincut-cutquery,AEGLMN22-star-contraction} to the
case of weighted graphs. These algorithms are randomized; the best
deterministic bound is $O(n^2/\log n)$~\cite{GrebinskiK00} which
allows full reconstruction of the input graph. There are also lower
bounds: \cite{GraurPRW20} define the {\em cut dimension} and use it to
give a lower bound of $3n/2-2$ queries for weighted graphs;
\cite{LeeLSZ21} improve the lower bound to $2n-2$, which remains the
best deterministic lower bound for this problem. The best randomized
lower bound for the problem is $\Omega(n\log \log n / \log n)$
queries~\cite{raz1995log,AssadiD21}.

There is earlier work on graph reconstruction using cut-value queries:
see, e.g.,
\cite{GrebinskiK00,bshouty2011parity,choi2013polynomial}. Recently,
there is work on solving other optimization problems (e.g., minimal
spanning forests) using cut value queries; see, e.g.,
\cite{harvey2008matchings,auza2021query,pmlr-v237-liao24b}. While we
do not consider these kinds of ``improper'' optimization problems in
this work. Further afield are more general query models, which allow
for ``independent set'' queries, see, e.g., \cite{pmlr-v237-liao24b}
for references.

\textbf{Continuous Optimization:}
There is a large body of work in \emph{continuous optimization} using comparisons between points, where the algorithm can query points $x,y \in \RR^d$ and get back the sign of $f(x) - f(y)$~\cite{JamiesonNR12,ChengSCC0H20,abs-2405-11454} (also see the work on \emph{dueling bandits}~\cite{YueBKJ12}). However, in our model we are allowed to only query discrete, feasible solutions and not points in the ambient space outside $\cF$, so the ideas used in gradient-based methods seem inapplicable. E.g., one can try writing each interior point as a convex combination of extreme points, but comparisons between these extreme points do not give us comparisons between the fractional points. Other approaches, e.g., estimating gradients or learning cutting-planes, also seem difficult to implement. For the special case of min-cuts, the Lov\'asz extension initially seems promising: we can learn the signs of the gradient coordinates, but this information is insufficient. Moreover, the Lov\'asz extension is not smooth; one can try Gaussian smoothing, but again that requires evaluation at points outside of our feasible set.

\textbf{Comparison Models more broadly:}
Ordinal preferences are considered in economics~\cite{pareto1919manuale}, recommendation systems and crowd-sourcing~\cite{chen2013pairwise}, modeling consumer
behavior~\cite{barnett2003modern}, statistics~\cite{bradley1952rank}, and the social sciences~\cite{heldsinger2010using}. More recently, these have become important to reinforcement learning with human feedback (RLHF), and particularly direct preference optimization methods~\cite{chen2025compopreferencealignmentcomparison,tang2024zerothorderoptimizationmeetshuman}.



\section{Linear Optimization for General Set Systems}
\label{sec:gen_set_system}

We begin our investigation with the question of query complexity for
linear optimization in a very general setting: in the \emph{general
  linear optimization} (GLO) problem, we are given a set of $N$ points
$\mathcal{P} = \{x_1, \dots, x_N\} \subseteq \mathbb{R}^d$, and an
unknown weight vector $w^* \in \RR^d$. We are only allowed comparison
queries: given $x, y \in \cP$, the comparison oracle responds with
$\operatorname{sign}\left( \langle w^*, x - y \rangle \right)$, i.e.,
the relative ordering of $x$ and $y$ in the ordering on $\cP$ induced
by $w^*$. The goal is to identify
$x^* := \arg\min_{x \in \mathcal{P}} \langle w^*, x \rangle$ using as
few queries as possible. The combinatorial optimization setting is
obtained when we restrict to some point set in $\{0,1\}^d$.

Our main result for general linear optimization depends on the
complexity of these point sets, via the notion of the \emph{\CDfull},
which we define next.

\subsection{Envelopes and the \CDfull of Point Sets}

We begin by defining the \emph{envelope} and \emph{\CDfull} of a point
set. Recall that a cone of a point set is the collection of all
positive linear combinations of these points. The idea behind the
envelope is simple: if $\sigma$ is the total order induced by $w^*$ on
the points $\{y_1, \ldots, y_n\}$ (i.e., if
$\ip{w^*, y_i - y_j} \geq 0$ for $i < j$, then taking the envelope of
$\sigma$ gives us all the ``implications'' of this ordering, and
allows us to derive implications without performing any further
calculations.

\begin{defn}
For a sequence of points $\sigma = (y_1, \dots, y_k) \in (\RR^d)^k$:
\begin{enumerate}[nosep]
    \item The \emph{envelope} of $\sigma$ is
    \[
        \envelope(\sigma) := \cone(\{y_j - y_i\}_{1 \le i < j \le k})
        = \cone(\{y_{i+1} - y_i\}_{1 \le i < k}).
    \]
    \item The points are \emph{conically independent} if, for each $t \in \{2, \ldots, k\}$,
    \[
        y_t - y_1 \notin \envelope(\sigma^{1:t-1}),
    \]
    where $\sigma^{1:t}$ denotes the prefix consisting of the first $t$ elements of $\sigma$.
\end{enumerate}
\end{defn}

\begin{defn}
  \label{defn:conic-dimension}
  The \emph{\CDfull} of a set $\cY \subseteq \RR^d$, denoted
  $\CDmath(\cY)$, is the largest integer $k$ for which there exists a
  length-$k$ conically independent sequence
  $\sigma = (y_1, \dots, y_k) \in \cY^k$.
\end{defn}
In other words, the \CDfull is the largest number of comparisons
$\ip{w^*, y_t}$ vs.\ $\ip{w^*, y_1}$ such that each one cannot be
inferred from the complete sorted ordering of the preceding points.

	



\subsection{An Algorithm for Finding a Minimizer}
\label{sec:geom-algo}

Given these definitions, it is possible to show that if a point set
$\cP$ has \CDshort $k$, then for the optimal point $y^* \in \cP$ there
exists proofs/certificates of optimality of size at most $k-1$ (see
\Cref{sec:certification}). Since we are interested in the query
complexity and not the certificate complexity of optimization, we show
the following:

\GLO*

The proof of this remarkable result relies on adapting the powerful
active classification framework of
\cite{KLMZ17-comp-based-classification} to the optimization setting;
we believe that this connection between the active
learning/classification and optimization settings should be
interesting more generally. The algorithm is also surprisingly simple:

\begin{algorithm}[H]
  \caption{Iterative Sieving Algorithm}
  \label{alg:sample-remove-recurse}
  \While{$|\cP|$ is more than $O(k)$}{
    \label{item:alg-step-1} Independently include each point of
    $\cP$ in a subset $\cY$ with probability $2k/N$ \\
    $\sigma \gets$ result of sorting $\cY$ using $O(k \log k)$ expected comparisons
    \\
    $y \gets$ first (smallest) element of $\sigma$ \\
    \label{item:alg-step-2} Eliminate all points
    $x \in \cP \setminus \{y\}$ such that
    $x - y \in \envelope(\sigma)$
  }
  \label{item:alg-step-3} Sort $\cP$ using a brute-force algorithm.
\end{algorithm}

The proof of the algorithm uses from the next lemma, which shows that
each iteration reduces the number of points in $\cP$ by half in
expectation. We defer the proof to \Cref{sec:proof-alg-progress}.

\begin{restatable}[Sieving Lemma]{lemma}{Sieve}
  \label{lem:alg-progress}
  The expected number of points that are eliminated from $\cP$ at
  step~\ref{item:alg-step-2} is at least $|\cP|/2$.
\end{restatable}

Since Boolean point sets have small \CDfull
\cite{KLMZ17-comp-based-classification} (see \Cref{sec:boolean-points}
for a proof), we get our main query complexity theorem for Boolean
optimization:

\BLO*

As mentioned in \Cref{sec:our-results}, this result is very surprising to
us: it shows that a small number of very structured comparison queries suffice
to find the optimal solution, even for NP-hard optimization problems for which the computational complexity is exponential (assuming
the ETH).

Can we hope to implement
\Cref{alg:sample-remove-recurse} efficiently for ``nice'' Boolean
problems? E.g., can we do this for problems which admit an optimization oracle, and
also a sampling oracle, so that we can implement
step~\ref{item:alg-step-1} efficiently? Matroids are one such
class of problems to keep in mind. Note that the above algorithm restricts to some
subset of the points in
step~\ref{item:alg-step-2}; it remains unclear how to efficiently sample
from these restrictions. To sidestep this difficulty, we give a different algorithm in the next section, and show that which we can indeed implement for several ``nice'' problems.

\section{Optimization from Certification: Dual Ellipsoid Algorithm}
\label{sec:ellipsoid}

In this section, we present an alternative to \Cref{alg:sample-remove-recurse} for the General Linear Optimization problem. Crucially, this framework establishes an efficient reduction from optimization to certification. The algorithm iteratively either acquires more information regarding the unknown objective vector $w^*$, or certifies that a candidate point $x\in \cP$ is optimal. 

This approach leverages the classic ellipsoid method using the concept of \textit{conic certificates} to determine if the candidate solution $x_t = \argmin_{x\in \cP} \langle w_{t-1}, x \rangle$ (which is optimal with respect to the current ellipsoid center $w_{t-1}$) is also optimal with respect to the hidden vector $w^*$ using only comparisons between feasible sets. The procedure is detailed in \Cref{alg:Dual-Ellipsoid}; we build up the necessary machinery for it in the next two sections.

\subsection{The Conic-Certification Oracle}
\label{sec:conic-cert-oracle}

The Dual Ellipsoid algorithm relies on having access to a specific
oracle, which we call the \emph{Conic-Certification Oracle} (CCO).  To
describe it, define the following notation: for any point
$y\in \RR^d$, let $\langle y\rangle$ denote its bit complexity. For a
subset $\cP\subseteq \RR^d$, let
$\langle \cP\rangle:= \max_{y\in \cP}\langle y\rangle$ be the maximum
bit complexity of all its points. 

\begin{algorithm}[H]
  \caption{Conic-Certification Oracle}
  \label{alg:CCO}
  \textbf{Input:} A vector $w \in \RR^d$. \\
  \textbf{Output:} A sequence $(y_1, \dots, y_k) \in \cP^k$ of points
  in $\cP$ satisfying:
  \begin{enumerate}
  \item \label{item:CCO1} $y_1 \in \argmin_{y\in \cP} \langle w, y \rangle$.
  \item \label{item:CCO2} $\langle w, y_1 \rangle \leq  \langle w, y_2 \rangle \leq \dots \leq \langle w, y_k \rangle$.
  \item \label{item:CCO3} $\cP \subseteq y_1+\envelope(y_1, \dots, y_k)$ and $k = \poly(d,\langle \cP\rangle)$.
  \end{enumerate}
\end{algorithm}

The first two conditions are natural: we output a sequence of points
such that $y_1$ is the minimizer, and each point has at least as much
cost as its predecessor. 
Condition~\ref{item:CCO3} requires some explanation: 
this
condition is equivalent to the statement that
\begin{align}
  \label{eqn:CCO-third-condition} \left(\langle w', y_1 \rangle \leq
  \langle w', y_2 \rangle \leq  \dots \leq \langle w', y_k
  \rangle\right) \implies \left(\langle w', y_1\rangle \leq \langle
  w', x\rangle, \quad \forall x\in \cP\right) 
\end{align}
holds for all $w' \in \RR^d$. In other words, the sequence
$(y_1, \dots, y_k)$ serves as a certificate: conditioned on the
inclusion $\cP\subseteq y_1 + \envelope(y_1, \dots, y_k)$, the
optimality of $y_1$ with respect to a weight vector $w$ is guaranteed
if the weights of the points in the sequence are
non-decreasing. (\Cref{sec:certification} gives more context, via
\Cref{lem:y-certif-cone-eqiv}.)

At a high level, the second and third conditions in \Cref{alg:CCO} can
be viewed as a valid certification of the optimality of $y_1$ (for the
input weight $w$) over all points in $\cP$, based solely on
comparisons within a small set of points $\{y_1, \dots, y_k\}$.

\subsection{The Ellipsoid Algorithm}
\label{sec:ellipsoid-algorithm}

Before describing our Dual Ellipsoid approach, we recall the standard
ellipsoid subroutine for rational polytopes in
\Cref{alg:ellipsoid-polyhedron}:

\begin{algorithm}[H]
  \caption{Ellipsoid Method}
  \label{alg:ellipsoid-polyhedron}
  \textbf{Input:}
  \begin{enumerate}
  \item A strong separation oracle $\mathrm{SEP}_P$ for a bounded polyhedron $P \subseteq \mathbb{R}^d$.
  \item A bit complexity upper bound $\phi$ for the inequalities defining $P$. \\
    \textit{(Note: This implies $P \subseteq B_2(\mathbf{0}, R)$ with $R=2^{O(d\phi)}$. Furthermore, if $P$ is full-dimensional, then $\vol(P)\geq 2^{-O(d^2\phi)}$.)}
  \end{enumerate}
  \textbf{Output:} A point $x \in P$, or correct assertion that $P$ is empty.
\end{algorithm}

For any query point $x \in \mathbb{Q}^d$, the separation oracle $\mathrm{SEP}_P$ asserts that $x \in P$ or returns a separating hyperplane (defined by a violated constraint of $P$) that separates $x$ from $P$.
\begin{theorem}[{\cite[Chapter 14]{Schrijver-LP-IP}, \cite[Chapter 3]{GrotschelLS88}}]
	\label{thm:ellipsoid-polyhedron}
	Let $P \subseteq \mathbb{R}^d$ be a full-dimensional rational polytope with facet complexity $\phi$. The Ellipsoid Method returns a point in $P$ or determines $P = \emptyset$ within the following bounds:
	
	\begin{enumerate}
		\item \textbf{Iterations:} The algorithm terminates after at most $T_E = O(d^3 \phi)$ calls to the separation oracle. This follows from $T_E = O(d \ln(\vol(B_2(\mathbf{0}, R))/\vol(P)))$ and the bounds on $R, \vol(P)$.	
		\item \textbf{Precision:} Calculations are performed with a precision of $L= O(T_E)$ bits after the binary point to ensure the ellipsoid contains $P$.		
		\item \textbf{Running Time:} The total time complexity is
			$O\left( T_E \cdot (d^2 \cdot \mathcal{M}(L) + T_{\mathrm{SEP}}) \right)$ where $\mathcal{M}(L)$ is the cost of arithmetic on $L$-bit numbers and $T_{\mathrm{SEP}}$ is the oracle cost.
	\end{enumerate}
\end{theorem}

\begin{remark}
  \label{rem:interior-ellipsoid}
  Assume that $P$ is guaranteed to be full-dimensional whenever it is
  non-empty. Furthermore, suppose the separation oracle is
  strengthened to report tight constraints for points on the boundary
  of $P$, rather than only reporting violated constraints (i.e., if
  $x \in \partial P$, the oracle returns a defining hyperplane). Under
  these conditions, the Ellipsoid algorithm in
  \Cref{alg:ellipsoid-polyhedron} returns a point in the \emph{strict
    interior} of $P$ or correctly asserts that $P$ is
  empty. 
\end{remark}

The final step before describing the dual ellipsoid algorithm is
defining the polyhedron $P$ we will work with and an implementation of
the strong separation oracle $\mathrm{SEP}_P$ using the
Conic-Certification Oracle.

\subsection{The Hyperplane Arrangement Induced by $\cP$}

Given a point set $\cP=\{x_1, \dots, x_N\}\subseteq \RR^d$, denote
\begin{align}
  H_{i,j}=\{w\in\RR^d: \langle w, x_i\rangle \geq \langle w, x_j\rangle \}, \text{and } h_{\{i,j\}}= \{w\in \RR^d: \langle w, x_i-x_j\rangle=0\}.
\end{align}
Notice that querying
$ \operatorname{sign}\left(\langle w^*, x_i - x_j \rangle\right) $ is
equivalent to checking whether $ w^* \in H_{i,j} $. The
\emph{hyperplane arrangement}
$ \mathcal{A} = \bigcup_{i,j \in [N]} h_{\{i,j\}} $ partitions
$ \mathbb{R}^d $ into \emph{cells}, where each cell is a maximal
connected region of $ \mathbb{R}^d $ that does not share any interior
points with any hyperplane in $ \mathcal{A} $. It follows that these
cells are convex polyhedra with facets defined by a subset of
hyperplanes from the arrangement.

\begin{defn}
  For each permutation $\pi \in S_N$, define
  \begin{align}
    C_{\pi}&=\{w\in \RR^d: \langle w, x_{\pi(1)}\rangle \leq \langle w, x_{\pi(2)}\rangle \leq  \dots \leq \langle w, x_{\pi(N)}\rangle \}\\
           &=\bigcap_{1\leq i\leq N-1}H_{\pi(i+1), \pi(i)}.
  \end{align}
  Similarly, for any partial order $\sigma$ on the set of indices $[N]$, define
  \begin{align}
    C_\sigma &= \{w \in \RR^d : \langle w, x_i \rangle \leq \langle w, x_j \rangle \text{ for all } i \preceq_\sigma j \}.
  \end{align}
\end{defn}

\begin{lemma}
  \label{lem:arrangement-cells}
  The following statements are true:
  \begin{enumerate}
  \item Any full dimensional $C_\pi$ is a cell of the arrangement $\cA$. Moreover, every cell of $\cA$ is a full dimensional set of the form $C_\pi$ for some $\pi \in S_N$.
  \item For any partial order $\sigma$, the set $C_\sigma$ is a convex polyhedron. Furthermore, $C_\sigma = \bigcup_{\pi \in \mathcal{L}(\sigma)} C_\pi$, where $\mathcal{L}(\sigma)$ denotes the set of linear extensions of $\sigma$ (permutations consistent with $\sigma$).
  \end{enumerate}
\end{lemma}

We defer the proof to~\Cref{sec: ellipsoid-omitted-proofs}. Define $\sigma^*$ to be the partial order on $[N]$ determined by the strict inequalities of $w^*$, such that $i \preceq_{\sigma^*} j$ if and only if $\langle w^*, x_j - x_i \rangle > 0$.

\begin{lemma}
	\label{lem:sigma^*}
	The weight vector $w^*$ lies in the interior of $C_{\sigma^*}$. Furthermore, for any $\widehat{w} \in \operatorname{int}(C_{\sigma^*})$, the implication
	\[ \langle \widehat{w}, x_i-x_j \rangle \leq 0 \implies \langle w^*, x_i-x_j \rangle \leq 0 \]
	holds for all $i, j \in [N]$.
\end{lemma}

\begin{cor}
	\label{cor:CCO-compatible}
	For any $\widehat{w} \in \operatorname{int}(C_{\sigma^*})$, if $\mathrm{CCO}(\widehat{w})=(y_1, \dots, y_k)$, then $\langle w^*, y_i - y_j \rangle \leq 0$ for all $1\leq i\leq j \leq k$. Consequently, $y_1 \in \operatorname*{argmin}_{x\in \cP}\langle w^*, x \rangle$.
\end{cor}

(Again, we defer these proofs to~\Cref{sec: ellipsoid-omitted-proofs}.) From \Cref{cor:CCO-compatible}, it is evident that identifying any point in the interior of $C_{\sigma^*}$ suffices to identify the global optimizer.
Naturally, we will set the polytope to $C_{\sigma^*}$ with an added box constraint $P:=C_{\sigma^*}\cap B_\infty(0,1)$.
We now describe how the CCO oracle can act as a separation oracle for $P$ as long as the optimal solution has not been found.

\subsection{CCO as a Separation Oracle}
\label{sec:cco-as-separation}

We implement the separation oracle for $P \subseteq C_{\sigma^*}$ using \Cref{cor:CCO-compatible}. Given a query point $\widehat{w}$ (the center of the current ellipsoid), we compute the sequence $(y_1, \dots, y_k) = \mathrm{CCO}(\widehat{w})$. We then check if this sequence is non-decreasing with respect to the hidden vector $w^*$ using the comparison oracle.

If the sequence is non-decreasing with respect to $w^*$, then by the properties of the CCO (specifically Equation \eqref{eqn:CCO-third-condition}), $y_1$ is an optimal solution, and the algorithm terminates.

Conversely, if the sequence is \emph{not} non-decreasing with respect to $w^*$, then $\widehat{w}$ cannot be in the interior of $C_{\sigma^*}$. In this case, there must exist a pair of indices $i, j$ such that the CCO ordering dictates $\langle \widehat{w}, y_i - y_j \rangle \leq 0$, but the true objective satisfies $\langle w^*, y_i - y_j \rangle > 0$. By the definition of $C_{\sigma^*}$, the condition $\langle w^*, y_i - y_j \rangle > 0$ implies that every point $w \in C_{\sigma^*}$ must satisfy $\langle w, y_i - y_j \rangle \geq 0$. Since $\widehat{w}$ violates this inequality, the hyperplane $\mathcal{H} = \{w : \langle w, y_i - y_j \rangle = 0\}$ validly separates $\widehat{w}$ from the target polytope $P$. 

We now have all the components for the dual ellipsoid algorithm, which
is presented as~\Cref{alg:Dual-Ellipsoid}:

\begin{algorithm}[H]
	\caption{Dual Ellipsoid Algorithm}
	\label{alg:Dual-Ellipsoid}
	\textbf{Initialization:} Bounding radius $R=\sqrt{d}$ for the initial elliposoid $\cE_0 = B_2(\mathbf{0}, R)$ and facet complexity $\phi= O(\langle \cP \rangle) $ as input to \Cref{alg:ellipsoid-polyhedron}.\\
	\textbf{For} $t = 1$ to $T$ \textbf{do}:
	
	\quad Let $(y_1, \dots, y_k) = \mathrm{CCO}(w_{t-1})$ be the conic certificate obtained for weight $w_{t-1}$.
	
	\quad Using the comparison oracle for $w^*$, verify if:
	\begin{align}
		\label{alg-line:comparison} \langle w^*, y_1 \rangle \leq  \langle w^*, y_2 \rangle \leq \dots \leq \langle w^*, y_k \rangle.
	\end{align}
	
	\quad \textbf{If} all comparisons are true \textbf{then}:
	
	\quad \quad \textbf{Return} $(\mathrm{YES}, y_1)$. \quad (\textit{$y_1$ is guaranteed to be an optimal solution})
	
	\quad \textbf{Else}: (\textit{a violation is found})
	
	\quad \quad Find indices $j, j+1$ such that $\langle w^*, y_j - y_{j+1} \rangle > 0$.
	
	\quad \quad Set normal $n_t = y_j - y_{j+1}$ and half-space $\cH_t = \{w \in \RR^d : \langle w, n_t \rangle \geq 0\}$. 
	
	\quad \quad \textit{(Note: $\cH_t$ separates $w_{t-1}$ from $w^*$)}.
	
	\quad \quad Set $w_t$ as the center of ellipsoid $\cE_t$ updated according to the seperating hyperplane $\cH_t$ by  \Cref{alg:ellipsoid-polyhedron}.
	
	\textbf{End For}
	
	\textbf{Return} $\mathrm{NO}$.
\end{algorithm}

\begin{theorem}
	\label{thm:dual-ellipsoid}
	\Cref{alg:Dual-Ellipsoid} is guaranteed to return $(\mathrm{YES}, y)$ with $y\in \argmin_{x\in \cP}\langle w^*, x\rangle$ for $T=O(d^3 \langle \cP \rangle )$.
\end{theorem}
\begin{proof}
The algorithm terminates in one of two ways. If the loop returns $(\mathrm{YES}, y_1)$ at step $t$, then by the properties of the CCO (specifically Equation \eqref{eqn:CCO-third-condition}), $y_1$ is an optimal solution. 

Conversely, if the algorithm returns $\mathrm{NO}$, the ellipsoid algorithm runs for all $T$ steps. At every step $t$, we construct a half-space $\cH_t$ derived from the violation $n_t = y_j - y_{j+1}$. Since $\mathrm{CCO}(w_{t-1})$ guarantees $\langle w_{t-1}, y_j \rangle \leq \langle w_{t-1}, y_{j+1} \rangle$, we have $\langle w_{t-1}, n_t \rangle \leq 0$, while the violation implies $\langle w^*, n_t \rangle > 0$. Thus, $\cH_t$ effectively separates the center $w_{t-1}$ from $w^*$. 

By \Cref{thm:ellipsoid-polyhedron}, such a sequence of separating cuts can have length at most $d^3\phi= O(d^3 \langle \cP\rangle )$ where $\phi$ is the bit-complexity of the seperating hyperplanes which we know is defined by normals of the form $y_i-y_j$ for $y_i, y_j \in \cP$. Therefore, for a sufficiently large $T=\Theta(d^3 \langle \cP \rangle )$, the algorithm must terminate with an optimal solution.
	\end{proof}

\subsection{Query Complexity and Running Time} 
\label{sec:query-compl-runn}

Let $T_D:= O(d^3 \langle \cP \rangle )$ be the number of iterations
given by \Cref{thm:dual-ellipsoid} that guarantees that
\Cref{alg:Dual-Ellipsoid} terminates with an optimal solution.
\begin{lemma}[Query complexity]
  \label{lem:dual-ellipsoid-qc}
  \Cref{alg:CCO} can be implemented (inefficiently) to output a
  sequence of length at most $\CDmath(\cP)$. The (inefficient) query
  complexity of \Cref{alg:Dual-Ellipsoid} is at most
  $T_D \cdot \CDmath(\cP)$.
\end{lemma}
\begin{proof}
  For an input $w$, consider the sequence $\sigma=(x_1, \dots, x_N)$
  of all points in $\cP$ in increasing order of
  $\langle w, x\rangle $. Let $(y_1=x_1, y_2 \dots, y_k)$ be the basic
  subsequence $B(\sigma)$. From \Cref{lem:basis-cone-property}, we
  know that
  $\cone(\{x_{i+1}-x_i\}_{1\leq i\leq N-1})\subseteq
  \envelope(B(\sigma))$. Using the fact that $y_1=x_1$, and
  $\cP \subseteq x_1+\cone(\{x_{i+1}-x_i\}_{1\leq i\leq N-1})$, we get
  $\cP\subseteq y_1+ \envelope(y_1, \dots, y_k)$ for
  $k\leq \CDmath(\cP)$. This sequence satisfies all the conditions
  required by \Cref{alg:CCO}.
	
	In step $t\in [T]$ of \Cref{alg:Dual-Ellipsoid}, the only place where the comparison oracle is used is to check \Cref{alg-line:comparison}. The number of comparisons is $k_t-1$, where $k_t$ is the size of the sequence output by $\mathrm{CCO}(w_{t-1})$. Hence, the query complexity is at most $\sum_{t=1}^T k_t$. Using the above implementation for $\mathrm{CCO}$, we know that each $k_t$ is at most $\CDmath(\cP)$, giving an overall upper bound of $T\cdot \CDmath(\cP)$.
\end{proof}
\begin{cor}
	\label{cor:DE-hypercube-qc}
	When $\cP\subseteq \{0,1\}^n$, the (inefficient) query complexity of \Cref{alg:Dual-Ellipsoid} is $\tO(d^5)$. 
\end{cor}
\begin{proof}
	Using $\langle \cP\rangle= O(d)$ and $\CDmath(\cP)=\tO(d)$ when $\cP\subseteq \{0,1\}^d$ (see \Cref{lem:conic-dim-boolean-upperbound}) in \Cref{lem:dual-ellipsoid-qc} yields the result.
\end{proof}
\begin{lemma}[Time complexity]
	\label{lem:dual-ellipsoid-tc}
	Given an implementation of $\mathrm{CCO}$ with time complexity $T_{CCO}$ for inputs $w$, \Cref{alg:Dual-Ellipsoid} can be implemented to run in $T_D\cdot (T_{CCO}+d^2\cdot \cM(d^3 \langle \cP \rangle )) $ time. Here, $\mathcal{M}(L)$ is the cost of arithmetic on $L$-bit numbers 
\end{lemma}
\begin{proof}
	Note that step $t$ of \Cref{alg:Dual-Ellipsoid}, makes one call to $\mathrm{CCO}$ with argument $w_t$  taking $T_{CCO}$ time and at most one Ellipsoid update taking $d^2\cdot \cM(d^3 \langle \cP \rangle )$ time. 
\end{proof}

\subsection{Application: Perfect Matchings in Complete Bipartite Graphs}
\label{sec:ellipsoid-applications}

In this section, we present an efficient algorithm for Boolean Linear
Optimization, when $\cP$ is the set of \emph{perfect} matchings in
complete bipartite graphs. Note that while
\Cref{sec:matroid-intersection} shows how to optimize over matchings
using comparisons, it allows us to compare matchings of all sizes. We
now show how to optimize over \emph{perfect} matchings by only
comparing \emph{perfect} matching---at least for complete bipartite
graphs. We achieve this by providing efficient implementations of the Conic-Certification Oracle (CCO). By \Cref{lem:dual-ellipsoid-tc}, these implementations suffice to ensure the efficient execution of \Cref{alg:Dual-Ellipsoid}.

Let $\cP$ be the set of characteristic vectors of perfect matchings in the complete bipartite graph $K_{n,n}$ (with $d=n^2$). Given an input weight vector $w$ for the CCO, the minimum weight perfect matching $y_1=\argmin_{y\in \cP}\langle w, y\rangle$ can be identified in $\poly(n, \langle w\rangle)$ time using standard bipartite matching algorithms.

It remains to generate a certificate of the optimality of $y_1$ based
solely on comparisons between perfect matchings in $\cP$. Without loss
of generality, assume $y_1$ consists of the edges
$\{(i,i) : i\in [n]\}$. Geometrically, the optimality of $y_1$ with
respect to $w$ corresponds to the existence of a halfspace $\cH$ with
normal $w$ that contains $\cP$, such that $y_1$ lies on the boundary
of $\cH$. By LP duality, such a halfspace can be represented as a
conic combination of the inequalities defining the polyhedron
$\mathrm{conv}(\cP)$. Our goal is to construct this certificate using
a small number of comparisons. The standard LP relaxation for the
bipartite perfect matching problem is:
\begin{align}
\min \Big\{ \sum_{i,j\in [n]}w_{i,j}x_{i,j} \mid 
    \sum_{i\in [n]}x_{i,j}= \sum_{j\in [n]}x_{i,j}=1, x_{i,j}\geq 0
  \Big\}, \tag{primal-PM}   \label{eqn:primal-PM} 
\end{align}
and its dual LP is:
\begin{align}
\max \Big\{ \sum_{i\in [n]}\alpha_i
                                +\sumL_{j\in [n]}\beta_j \mid
                                \alpha_i +\beta_j \leq w_{i,j} \quad
                                \forall i,j \in [n]
  \Big\}. \tag{dual-PM}      \label{eqn:dual-PM}  
\end{align}

Proving the optimality of $y_1$ with respect to $w$ is equivalent to demonstrating the feasibility of the following system:
\begin{align}
\alpha_i +\beta_i &=w_{i,i} \quad \forall i \in [n]\\
\alpha_i +\beta_j &\leq w_{i,j} \quad \forall i,j \in [n].
\end{align}

Eliminating the variables $\beta$, this feasibility problem simplifies to:
\begin{align}
\alpha_i -\alpha_j &\leq w_{i,j}-w_{j,j} \quad \forall i,j \in [n].
\end{align}
This condition is equivalent to requiring that the complete directed graph $D$ on vertex set $[n]$ with edge weights $c_{i,j}:= w_{i,j}-w_{j,j}$ contains no negative cycles. (Note that the optimality of $y_1$ guarantees that $D$ has no negative cycles; our task is to certify this fact.) Observe that the weight of any directed cycle $C$ (or more generally, a vertex-disjoint union of cycles, i.e., a 2-factor) in $D$ corresponds exactly to the weight difference between two perfect matchings.

Thus, the problem reduces to the following: Given a complete directed graph known to have no negative cycles, certify this property by comparing the weights of 2-factors. Note that since the empty 2-factor has weight zero, we may also check the sign of any 2-factor.

We achieve this by running the Bellman-Ford algorithm from each vertex $s\in [n]$. Let $DP_{s}(v, k)$ denote the shortest path from $s$ to $v$ using at most $k$ edges. We initialize $DP_{s}(s, 0)$ as the empty path and $DP_{s}(u,1)$ as the edge $(s, u)$ for $u\neq s$. The entry $DP_s(v, k+1)$ is updated by selecting the minimum weight path from the set of candidates:
\begin{align}
 \{DP_s(u, k)\cdot (u, v)\}_{u\in [n]}\cup \{DP_s(v, k)\}.
\end{align}
Note that a candidate path $DP_s(u, k)\cdot (u, v)$ might visit $v$ more than once. However, such walks must have weight strictly greater than $DP_s(v, k)$; otherwise, the graph would contain a negative cycle involving $u$ and $v$ (which we can rule out by checking the sign of this cycle). After this filtering, all candidate sets consist of simple $s-v$ paths. The minimum weight path among them can be identified by comparing the cycles formed by adding the edge $(v, s)$ to each candidate path. 

Finally, if the DP completes without identifying any negative cycles
during the filtering steps, it correctly computes the shortest paths
from $s$ to every vertex $v\neq s$. Verifying that none of the cycles
$DP_s(v, n-1)\cdot(v,s)$ have negative weight confirms that $D$
contains no negative cycles, thereby certifying the optimality of
$y_1$.


\section{Global Subspace Learning}
\label{sec:GSL}
In this section, we develop our second general algorithmic framework, tailored to cases where the weight vector $w^*$ in the objective function uses integers bounded by $B$, i.e., $w^*\in \ZZ^n$ and $\|w^*\|_{\infty}\leq B$. The resulting algorithm has a clean linear-algebraic structure which we can use to obtain efficient algorithms for several natural set systems $\cF$.

Recall that the iterative sieving algorithm from \Cref{sec:gen_set_system} was based on iteratively sampling and sorting to filter out dominated solutions, while the Dual Ellipsoid algorithm in \Cref{sec:ellipsoid} maintained a geometric region of plausible weights to either certify optimality or shrink the dual space. Our new \emph{Global Subspace Learning (GSL)} algorithm takes a more direct algebraic approach. It iteratively learns the linear subspace of directions orthogonal to $w^*$. This algorithm is highly intuitive and demonstrates the idea of inference naturally. It also achieves a stronger global guarantee: it actually sorts \emph{all} feasible sets by their weight, using only $\poly(nB)$ comparisons---a task which is feasible in the bounded weights case, because there are at most $O(nB)$ possible distinct weight classes.
\subsection{The Algorithm}

As in the previous section, we work with a set $\cP$ of points, which are
indicator vectors $\mathbbm{1}_S$ of the feasible sets in $S\in
\cF$. For brevity, we will use $x,y \in \cP$ to refer both to points
in $\RR^n$, and also the corresponding sets in $\cF$. The weight of a
point $x$ is $w^*(x) := \langle w^*, x\rangle$.

The technical core of the algorithm is the following observation: if
two points $x, y\in \cP$ have the same weight, they imply the linear
equality $\langle w^*, x-y\rangle =0$. Concretely, the \emph{Global
  Subspace Learning} algorithm does the following:
\subsubsection{Global Subspace Learning Algorithm}
  \label{alg:GSL}  
  \begin{enumerate}
\item \textbf{Buckets:} At the end of step $t$, the algorithm has
  considered $t$ points $y_1, \ldots, y_t$ from $\cP$ which are divided into some number
  of buckets $B_1, B_2, \dots, B_{\ell}$. These buckets are indexed in
  increasing order of weight, and each bucket contains points of the
  same weight. For each bucket $B_i$, we have chosen an arbitrary
  \emph{representative} $r_i$, say the first point in that bucket.

  Moreover, the algorithm also maintains a subspace
  \[\cA = \textnormal{span}\left(\cup_{i \in [\ell]}\{x-y \mid x, y\in B_i\}\right)\]
  that is spanned by directions $x-y$ known to be orthogonal to
  $w^*$. By construction, if a pair $x', y'\in \cF$ satisfies $x'-y'\in \cA$, we can infer that
  $\ip{w^*, x'-y'} = 0$ and hence $w^*(x')=w^*(y')$ . Hence, each bucket $B_i$ has a
  corresponding affine subspace $\cA_i=\{r_i+ v \mid v\in \cA\}$
  defined using the representative $r_i \in B_i$, and $\cP \cap \cA_i$ is precisely the set of points which can be inferred to
  have the same weight as the points in $B_i$ at this time.

  To begin, at time $t = 1$, we choose $y_1$ as an arbitrary point in
  $\cP$, have a single bucket $B_1 = \{y_1\}$ and $r_1 = y_1$,
  and the subspace $\cA$ is the trivial subspace of dimension zero.
  
\item \textbf{Separation Step:} At step $t+1$, the algorithm selects a
  point $y_{t+1} \in \cP$ that does not belong to
  $\bigcup_{i\in [\ell]} \cA_i$ if such a point exists,
  else it terminates. In other words, it selects a point whose weight
  cannot be inferred to equal that of any representative $r_i$, based
  on the knowledge accumulated so far. (This is the central \emph{computational} challenge of the algorithm, which will return to in the next section.)

\item \textbf{Update Step:} It uses binary search to locate the
  position of the weight $w^*(y_{t+1})$ with respect to the ordered
  sequence $w^*(r_1) < \dots < w^*(r_\ell)$.

  If $w^*(y_{t+1})=w^*(r_i)$ for some $i\in [\ell]$, then it inserts
  $y_{t+1}$ into $B_i$, and updates $\cA$ to include the new direction
  $(y_{t+1}-r_i)$ in its span, thereby increasing its dimension. This
  also changes the definitions of all the affine subspaces
  $\cA_j$. Else the weight $w^*(y_{t+1})$ is a weight it has not seen
  before, so the algorithm starts a fresh singleton bucket containing
  $y_{t+1}$ as its representative.
\end{enumerate}

\begin{lemma}
  \label{lem:GSL-structural}
  The number of iterations of the Global Subspace Learning algorithm
  is at most $2nB+n$.
\end{lemma}
\begin{proof}
  The potential $\phi=\dim(\cA)+\ell$, starts at zero and is
  upper-bounded by $(n-1)+(2nB+1)=2nB+n$. It is sufficient to prove
  that $\phi$ increases by exactly $1$ each time a new point is
  encountered. We have two cases when a new point $y_{t+1}$ is considered:
  \begin{enumerate}
  \item If $w^*(y_{t+1})=w^*(r_i)$ for some $i\in [\ell]$, the number
    of buckets $\ell$ remains unchanged, as $y_{t+1}$ is inserted into
    $B_i$.

    This vector is guaranteed to be linearly independent of $\cA$ thus
    increasing the dimension because $y_{t+1}-r_i \notin \cA$, which
    follows directly from the condition $y_{t+1}\notin \cA_i$.
  \item Else, a new weight value is discovered. A new bucket is
    opened, which increases $\ell$ by one, while $\cA$ remains
    unchanged.
  \end{enumerate}
  In both cases, the potential function $\phi$ increases by exactly
  one. This completes the proof.
\end{proof}

\begin{restatable}[Boolean Linear Optimization: Bounded-Weights]{theorem}{bounded}
  \label{thm:bounded}
  For any family $\mathcal{F} \sse 2^U$ and unknown integer weight
  function satisfying $|w^*_e| \leq B$, we can sort the feasible sets
  according to their weight $w^*(S)$ and hence solve
  $\arg\min_{S\in \cF} w^*(S)$ using $O(nB \log nB)$ comparison
  queries, where $|U| = n$. In fact, the number of comparison queries
  required is $O((n+C)\log C)$ if $C$ is the number of distinct
  weight values realized by the feasible set.
\end{restatable}
\begin{proof}
  Algorithm \ref{alg:GSL} maintains a structured representation which allows us
  to identify the weight class of any point $x \in \cP$. We can decide that $w^*(x)$ is the $i$th smallest weight by checking if
  $x\in \cA_i$ (or equivalently, $x-r_i\in \cA$), simply because the
  algorithm's termination condition ensures that
  $\cP \subseteq \bigcup_{i\in [\ell]} \cA_i$. This serves as a
  compact representation for the sorted order of all points by their
  weights $w^*(x)$. Moreover, $r_1$ (or any other point in $B_1$) is a
  point of minimum weight.

  The number of comparison queries per point encountered is
  $O(\log \ell)=O(\log nB)$ due to binary search over at most $O(nB)$
  weights, summing up to $O(nB\log(nB))$ queries over its entire
  execution. This bound can be improved to $O((n+C)\log C)$ if $C$ is
  the number of distinct weight values realized by sets in $\cF$,
  because $O(\log \ell)=O(\log C)$ and the number of points
  encountered is at most $n-1+C$.
\end{proof}

The GSL framework provides a powerful information-theoretic bound. More importantly, it provides a clear path toward computationally efficient algorithms, which we turn to in the next section. We make two final observations:

\begin{enumerate}
\item A conceptual difference between
  \Cref{thm:query-comp-boolean} and \Cref{thm:bounded} is that the former uses
  {\em conic spans} while the second result uses only \emph{linear
    spans} for inference. The use of linear spans makes it easier to
  reason about, and as we will see, to obtain efficient algorithms.
  
\item The GSL algorithm can be implemented even when restricted to
  only performing equality queries---i.e., queries that only tell us
  whether $w^*(S)=w^*(T)$ or $w^*(S)\neq w^*(T)$. In this setting, the
  buckets would be unordered, and identifying whether
  $w^*(y_{t+1})=w^*(r_i)$ would require $O(nB)$ equality queries,
  increasing the query complexity to $O(n^2B^2)$. More generally, $O((n+C)\cdot C)$ equality queries if $C$ is the number of distinct weight values realized by the feasible set.
\end{enumerate}

\subsection{Efficient Implementation of Global Subspace Learning}
\label{sec:gsl-efficient}

The main computational bottleneck in the GSL algorithm is in finding a
point $y \in \cP$ such that
$y \notin\bigcup_{i \in [\ell]} \cA_i$---i.e., the separation problem
over the union of the affine subspaces $\cA_i$. We show that we
can solve this separation problem when $\cF$ is 
the family of all $k$-subsets $\binom{U}{k}$, or more
generally, when $\cF$ is the set of bases of a linear matroid
representable over $\RR$.

Recall that the subspace $\cA$ is the span of directions orthogonal to
$w^*$; since $w^*$ has integer coordinates, it must lie in the
orthogonal lattice $\cW := \cA^{\perp} \cap \ZZ^n$. Let
$A \in \ZZ^{m \times n}$, where $m=\dim(\cA)\leq n-1$, be the matrix
whose rows form a basis for $\cA$. By definition, the orthogonal
complement $\cA^\perp$ is the nullspace of $A$. Therefore, $\cW$ is
precisely the \emph{integer nullspace} of
$A$.

An integer basis for $\cW$ can be computed using the \emph{Smith
  Normal Form} (SNF) of the $m \times n$ matrix $A$, which returns a
decomposition $D = UAV$, where $U \in \mathbb{Z}^{m \times m}$ and
$V \in \mathbb{Z}^{n \times n}$ are unimodular matrices, and $D$ is an
$m \times n$ diagonal matrix with $m$ non-zero entries. The last
$\codim:=n-m$ columns of $V$, which correspond to the zero columns of $D$,
form the required integer basis for $\cW$. Moreover, the decomposition
$U, D, V$ can be found in polynomial time.

\begin{lemma}[Integer Nullspace via SNF, \cite{KannanBachem1979, Schrijver-LP-IP}]
  \label{lem:integer-nullspace}
  Let $A \in \{-1,0,1\}^{m\times n}$. An integer
  basis for the integer nullspace
  \[
    \ker_\ZZ(A) = \{x \in \ZZ^n : A x = 0\}
  \]
  can be computed in polynomial time in $n$ and $m$. The
  entries of the resulting basis vectors are bounded in bit-length by
  $\mathrm{poly}(n, m)$.
\end{lemma}

Note that our matrix entries are of the form $\mathbbm{1}_{S}-\mathbbm{1}_{T}$ where $S, T \in \cF$, and hence they lie in $\{-1,0,1\}$. Since
$m\leq n$, \Cref{lem:integer-nullspace} implies that the resulting
integer basis vectors have entries bounded by
$2^{\mathrm{poly}(n)}$. Let $W \in \ZZ^{\codim\times n}$ denote the matrix
whose rows form this basis spanning the orthogonal lattice $\cW$. The
next implication is immediate: 

\begin{obs}
  \label{obs:GSL}
  Point $x$ belongs to $\cA_i$ for some $i\in [\ell]$ if and only
  if $W(x-r_i)=0$. This implies that
  $y \notin \bigcup_{i \in [\ell]} \cA_i$ is equivalent to
  $Wy\notin \{Wr_1, \dots, Wr_\ell\}$, where the $r_i$ vectors are the
  bucket representatives.
\end{obs}

Let $W=[w_1, \dots, w_n]$ where $w_i \in \ZZ^s$ is the $i$th column of
$W$ and $Z=\{z_1, \dots, z_{\ell}\}$ with $z_i=Wr_i$. We will use
$w(S):=\sum_{i\in S}w_i$. Henceforth, the following problem is the
one we focus on.

\begin{defn}[Separation Problem]
\label{defn:separation}
  Given vectors $w_1, w_2, \ldots, w_n \in \ZZ^\codim$, and a
  collection $Z \sse \ZZ^\codim$ with some $\ell$ vectors, identify a
  subset $S\in \cF$ such that $w(S)\notin Z$.
\end{defn}

\subsubsection{Efficient Separation via Dynamic Programming}
\label{sec:gsl-hypercube}
As a quick warm-up, we show an efficient algorithm for the separation problem using dynamic programming, when the feasible set $\cF$ is $2^U$. (Note that
even these cases were not efficiently implementable using the Sieving
algorithm of \Cref{sec:gen_set_system}).

Define 
\[
\cV_i= \{w(S): S\subseteq [i]\},\quad 0\leq i\leq n,
\]
as the set of vector values obtained by adding a subset of the first $i$ columns of $W$. We can compute $\cV_i=\cV_{i-1}\cup \{w_i+v: v\in \cV_{i-1}\}$ starting from $\cV_0=\emptyset$ and increasing $i$ stopping when $|\cV_i|>\ell$. This implies that there exists a value $v\in \cV_i$ that is not in $Z$ as $|Z|=\ell$. 

The set $S$ such that $w(S)=v$ can be constructed by tracking the computation of $\cV_i$.

\subsubsection{Efficient Separation for Linear Matroids via Algebraic Techniques}
\label{sec:gsl-linear-matroid}

In this section, we show how to solve the separation problem for $\cF$
being the bases of a linear matroid $\cM = (U, \cI)$. Recall that a
linear matroid of rank $k$ is represented by vectors
$V = \{v_1, \dots, v_n\} \subset \RR^k$, where each $v_i$ corresponds
to an element in $U$. The independent sets $\cI$ are those subsets of
$U$ whose corresponding vectors are linearly independent, and the
bases $\cF$ are the independent sets of size $k$.

In order to solve the separation problem for vector costs $w_i$, we
proceed in two steps:
\begin{enumerate}
\item First, we solve the problem when the dimension of the vectors in
  the separation problem is $\codim = 1$, i.e., we are given a
  collection of scalars $c_1, \dots, c_n$, and a collection $Z$ of
  scalars, and want to find a feasible set $S\in \cF$ such that
  $\sum_{i \in S} c_i \notin Z$.
\item Secondly, we encode each vector $w_i$ into a unique non-negative
  scalar $\widehat{w}_i$ (and each vector in $Z$ also to a scalar),
  and apply the result from the first step.
\end{enumerate}

The scalar encoding in the second step is easy to describe. For a vector 
$v \in [-M,M]^\codim$, define its encoding 
\[ g(v) := (3nM)^\codim + \sum_{j=1}^\codim (3nM)^{j-1} v_{j}. \]


\begin{lemma}[Scalar Encoding]
\label{lem:scalar-encoding}
 For any vector $v \in [-M,M]$, its encoding $g(v)$ takes on values in $[0,(3nM)^{\codim +1}]$. Moreover, for any two subsets $S, S' \subseteq U$, the sums of the encoded values are equal if and
only if the sums of the original vectors are equal: i.e.,
$\sum_{i \in S} g(w_i) = \sum_{i \in S'} g(w_i)
 \iff 
\sum_{i \in S} w_i = \sum_{i \in S'} w_i$. 
\end{lemma}

Given this, we turn to the first step, and give an algorithm for the
separation problem when the columns $w_i$ of $W$ are scalars:
\begin{restatable}[Separation for Linear Matroids]{lemma}{LinSep}
  \label{lem:dif-weight-basis}
  Given a linear matroid $\cM=(U, \cI)$, non-negative integers
  $c_1, c_2, \dots, c_n$ for elements in $U$, and a set of non-negative values
  $Z=\{z_1, \dots, z_{\ell}\}$, we can identify a basis $S\in \cF$
  of $\cM$ such that $c(S)\notin Z$ (or decide that no such basis
  exists) in $\poly(n, \ell, \log M)$ time, where
  $M=\max_{i\in [n], j\in [\ell]}\{c_i, z_j\}$.
\end{restatable}

To give the main idea behind the lemma, let us first give a proof that
yields a running time of $\poly(n,M)$. The full proof showing a
$\poly(n, \log M)$ runtime is deferred to the appendix.

\begin{proof}[Warmup Proof of \Cref{lem:dif-weight-basis}]
  The main idea is to use a specific determinant as a generating
  function for the set of basis costs, $\{c(S): S\in \cF\}$.  For a
  linear matroid of rank $k$ represented by
  $V=\{v_1, \dots, v_n\}\subseteq \RR^k$, we use the Cauchy-Binet
  identity to infer:
  \[
    \det\bigg(\sum_{i=1}^n x_i v_i v_i^T\bigg)
    = \sum_{S \in \cF} \bigg( \prod_{i \in S} x_i \bigg) \det(V_S)^2,
  \]
  where $V_S$ is the $k \times k$ matrix with columns indexed by $S$,
  and $\det(V_S) \neq 0$ if and only if $S \in \cF$.
  We now define a polynomial $P(x)$ by setting $x_i = x^{c_i}$, where
  $c_i$'s are the given scalars: i.e.,
  \begin{gather}
    P(x) = \det\bigg(\sum_{i=1}^n x^{c_i} v_i v_i^T\bigg)
    = \sum_{S \in \cF} x^{\sum_{i \in S} c_i} \det(V_S)^2 = \sum_{S
      \in \cF} x^{c(S)} \det(V_S)^2. \label{eq:3}
  \end{gather}
  The exponents with non-zero coefficients in $P(x)$ are precisely the
  basis costs $\{c(S) \mid S \in \cF\}$. The degree of $P(x)$ is
  bounded by $nM$. We can therefore find $P(x)$ exactly using
  polynomial interpolation by evaluating it at $O(nM)$ points. This
  process runs in $\poly(n,M)$ time. Once the polynomial $P(x)$ is
  known, we can inspect its coefficients to find if any exponent
  $c \notin Z$ has a non-zero coefficient.

  Furthermore, the associated basis can be found using
  self-reducibility for matroids. For each element $j \in U$, we
  compute the ``edge-deletion'' polynomial
  $P_j(x)= \det(\sum_{i\neq j} x^{c_i} v_i v_i^T)$: if $c$ appears as
  an exponent in $P_j(x)$, a basis with cost $c$ exists without
  $j$. We delete $j$ and recurse on the smaller problem. On the other
  hand, if $c$ does not appear in $P_j(x)$, then $j$ must be in every
  basis $S$ with $c(S)=c$. We select $j$ and recurse, searching for a
  basis with cost $c-c_j$ in the problem on $U / \{j\}$ (i.e., using
  an ``edge-contraction'' polynomial $P_{/j}(x)$)
\end{proof}

When the bound $M$ is large, the $\poly(n,M)$ time bound is
infeasible, as the degree of $P(x)$ is too high for efficient
interpolation. The full proof, which appears in
\Cref{sec:GSL-appendix}, avoids this problem by remaindering; i.e.,
working with the polynomial $P(x) \pmod{(x^p-1)}$ for a suitable prime
$p \in \poly(n,\ell,\log M)$. This prime is chosen so that the residues
modulo $p$ of all values in $Z$ and at least one basis cost
$c \notin Z$ (if such a basis exists) are distinct. This allows for
efficient interpolation and reconstruction within the required time
bounds. Now we can use \Cref{lem:dif-weight-basis} to prove \Cref{thm:LinMatOpt}.

\LinMatOpt*
\begin{proof}
  From Algorithm~\Cref{alg:GSL}, the stated query complexity follows directly from \Cref{thm:bounded}. 
  It remains to bound the total running time required to implement the separation steps. 

  The maintenance of buckets, affine subspaces, and the corresponding updates can each be carried out in $\mathrm{poly}(n, B)$ time overall. 
  By \Cref{lem:GSL-structural}, the number of separation steps is bounded by $O(nB)$. 
  Each separation step invokes the procedures in \Cref{lem:scalar-encoding} and \Cref{lem:dif-weight-basis}, 
  which take $\mathrm{poly}(n, \log M)$ and $\mathrm{poly}(n, \ell, \log M)$ time, respectively. 
  Summing over all separation steps gives a total time of $\mathrm{poly}(n, B)$, 
  using the facts that $\ell = O(nB)$ and $M = 2^{\mathrm{poly}(n)}$, the latter following from \Cref{lem:integer-nullspace}.
\end{proof}

\subsection{Other Applications}
\label{sec:other-apps}

In this section, we present several applications of \Cref{thm:LinMatOpt} and \Cref{thm:bounded} that yield low-query and efficient algorithms for well-studied problems such as $k$-SUM, SUBSET-SUM, and sorting sumsets $A+B$ (see \cite{kane2019near}). When the elements of the ground set take integer values in the range $[-B, B]$, our algorithms achieve better query complexity while also being the first efficient algorithms in this setting.

\subsubsection{Low-query and efficient comparison decision trees}

\begin{enumerate}
    \item \textbf{$k$-SUM:}  
    In the $k$-SUM problem, the input elements in $U$ have unknown values $w_e^*$, and the goal is to decide whether there exist $k$ distinct elements whose sum is zero. The comparison model considered in \cite{kane2019near} allows comparisons between $w^*(S)$ and $w^*(T)$ for $S, T \in \binom{U}{k}$ and between $w^*(S)$ and $0$. They achieve a query complexity (decision tree depth) of $O(kn\log^2 n)$.
    
    Using \Cref{thm:LinMatOpt} with $C = kB$, we obtain an algorithm with query complexity $O((n + kB)\log(kB))$ and running time $\poly(n, B)$, which improves both measures when $B = o(n)$. Our algorithm first sorts all feasible $k$-sets by their weights using only comparison queries between solutions, and then performs a binary search using comparisons with $0$ to check whether a subset of size $k$ has total weight zero.

    \item \textbf{SUBSET-SUM:}  
    In the SUBSET-SUM problem, the goal is to determine whether there exists a subset $S \subseteq U$ such that $w^*(S) = t$ for a given target $t$. Similar to $k$-SUM, the model allows comparisons between feasible sets ($\cF = 2^U$) and between a feasible set and the target $t$. The algorithm in \cite{kane2019near} achieves query complexity $O(n^2\log n)$.
    
    Using Algorithm~\Cref{alg:GSL} with the efficient separation step stated in \Cref{sec:gsl-hypercube}, we obtain an algorithm with query complexity $O(nB \log(nB))$ and running time $\poly(n, B)$, which again improves both parameters when $B = o(n)$. As before, the algorithm first sorts all subset weights using comparison queries and then performs a binary search using comparisons with $t$ to determine whether a subset with total weight $t$ exists.

    \item \textbf{Sorting sumsets $A+B$:}  
    In the $A+B$ sorting problem, we are given two sets $A$ and $B$, each of cardinality $n$, with unknown element values $w^*_a$ and $w^*_b$ for $a \in A$ and $b \in B$. The goal is to sort the set $A+B = \{a+b : a \in A, b \in B\}$ using only comparisons between sums of the form $w^*(a_1) + w^*(b_1)$ and $w^*(a_2) + w^*(b_2)$.
    
    The model in \cite{kane2019near} even allows stronger comparisons—such as comparing sums of four elements $w^*(a_1) + w^*(b_1) + w^*(a_2) + w^*(b_2)$ versus $w^*(a_3) + w^*(b_3) + w^*(a_4) + w^*(b_4)$—which we do \emph{not} use. Their algorithm achieves a query complexity of $O(n \log^2 n)$.
    
    To fit this problem into our framework, we set $U = A \cup B$, and let $\cF$ consist of the bases of a partition matroid over $U$ with parts $A$ and $B$, each having capacity one. Then comparing $w^*(a_1) + w^*(b_1)$ with $w^*(a_2) + w^*(b_2)$ corresponds exactly to comparing the weights of two feasible sets $\{a_1, b_1\}$ and $\{a_2, b_2\}$. Since partition matroids are representable over the reals, and the element values lie in $[-B, B]$, we can apply \Cref{thm:LinMatOpt} with $C = 2B$ to obtain an algorithm with query complexity $O((n + B)\log B)$ and running time $\poly(n, B)$. This improves on previous bounds when $B = o(n)$ and uses only valid pairwise comparisons among elements of $A+B$.
\end{enumerate}
\medskip
\noindent
Moreover, our results extend to an even weaker model in which only \emph{equality} queries are allowed. In this setting, the query complexities for the three problems above are $O(kB(n + kB))$, $O(n^2B^2)$, and $O(B(n + B))$, respectively, while still maintaining a running time of $\poly(n, B)$. See \Cref{tab:applications} for a summary comparing the results.

\section{Minimum Cut using Cut Comparisons}
\label{sec:minimum-cut}

We now turn to specific combinatorial problems, and give efficient algorithms using their structural properties. We begin with the min-cut problem on graphs and prove \Cref{thm:unweighted-cuts,thm:graph_recovery}, which we restate for convenience.

\mincut*

\graphrecov*

In order to prove the two results, we draw on the following useful
primitives that can be obtained using the cut comparison queries:

\begin{restatable}[Structural Primitives]{theorem}{CutStructureLemma}
	\label{lem:cut-query-structural}
	When $n\geq 4$, for any vertex $u \in U$:
	\begin{enumerate}[label=(\roman*),itemsep=0.05em,topsep=0.05em]
		\item \label{item:struct1} Given a set of $k$ vertices
		$A = \{ v_1, \dots, v_k\} \sse U \setminus \{u\}$, the neighbors
		of $u$ in $A$ can be determined using $k + O(\log n)$ queries and $\tO(n+k)$ time.
		
		\item \label{item:struct2} Given an ordered subset
		$T = (v_1, \dots, v_t) \subseteq U \setminus \{u\}$, either the
		neighbor $v_i \in N(u)$ of $u$ with the smallest index $i \in [t]$
		(or a certificate that $N(u)\cap T=\emptyset$) can be found using
		$O(\log n)$ queries and $\tO(n)$ time.
	\end{enumerate}
\end{restatable}

\begin{remark}[A Note on Runtimes]
	\label{rem:runtime-basic}
	We assume that the query input $(S,T)$ to the comparison oracle is given as two bit vectors encoding the sets $S$ and $T$ respectively. The running time bounds are the total computation done to process the query information, where oracle calls are assumed to take unit time. 
	The dominant cost is writing down the input bit vectors for queries, which takes $O(n)$ time per query. In addition, we separately account for standard algorithmic overhead, such as binary search  or other data structure updates.
\end{remark}

\subsection{Graph Reconstruction}
\label{sec:graph-reconstruction}

With the power of \Cref{lem:cut-query-structural} behind us, the graph
recovery problem is easily solved. Indeed, using
\Cref{lem:cut-query-structural}\ref{item:struct1}, we can discover
the neighborhood of node $u \in V$ using $n-1+O(\log n)$ queries and $\tO(n)$ time;
summing over all nodes gives us the graph $G$ in at most $O(n^2)$
queries and $\tO(n^2)$ time.

To get the better bound of $O(m \log n)$ for sparse graphs, we use 
\Cref{lem:cut-query-structural}\ref{item:struct2} to prove:
\begin{restatable}[Edge Extraction]{lemma}{EdgeExtraction}
	\label{lem:edge-extraction}
	When $n\geq 4$, for any vertex $u$, a subset $T\subseteq U-u$ of vertices, and any
	integer $k$, we can   extract $r=\min(k,\partial(u,T))$ edges
	from $\partial(u,T)$ using $O(r\log n)$ queries and $\tO(n)$ time. 
\end{restatable}

\begin{proof}
	Start with $T_0:=T$ and extract a neighbor
	$v_i\in N(u)\cap T_i$ (if any) from $\partial(u,T_i)$ using
	\Cref{lem:cut-query-structural}\ref{item:struct2}; set
	$T_{i+1}:=T_i-v_i$, and repeat. Each step requires $O(\log n)$
	queries and $\tO(n)$ time. Repeating this for
	$r = \min(k, \partial(u,T))$ neighbors gives a total of
	$O(r \log n)$ queries and $\tO(nr)$ time. A more efficient
        implementation, which amortizes the search and is detailed in
        \Cref{app:cut-proofs}, achieves the claimed $\tO(n)$ time. 
\end{proof}

Given this efficient edge-extraction claim, we now prove the
claimed bounds for graph recovery.

\begin{proof}[Proof of \Cref{thm:graph_recovery}]
	When $n\geq 4$,
	using \Cref{lem:edge-extraction}, we can discover
	all the neighbors of $u$ using $|\partial(u)|\cdot O(\log n)$
	queries and $\tO(n)$ time. Summing over all $u$ gives a total of at most $O((m+n) \log n)$
	queries and $\tO(n^2)$ time. Hence, we start by running this procedure until the number
	of edges discovered is at least $n^2/\log n$. At this point, we
	switch to the algorithm using
	\Cref{lem:cut-query-structural}\ref{item:struct1} to discover the
	remaining edges using at most $O(n^2)$ queries. This guarantees that
	we use at most $O(\min((m+n)\log n,n^2))$ queries and $\tO(n^2)$ time.
	
	When $n\leq 3$ and $G \notin \{K_2, \bar{K}_2, K_3, \bar{K}_3\}$, sorting the degrees of the vertices reveals the graph.
\end{proof}

\subsection{Finding Minimum Cuts}
\label{sec:cuts-algos}
We now turn to finding minimum cuts. We start off by observing that
when $n\leq 4$, we can bruteforce the minimum cut and otherwise, use \Cref{thm:graph_recovery} to optimize over min-cuts:
we reconstruct the graph using $O(\min((m+n)\log n,n^2))$ queries and $\tO(n^2)$ time; we can then optimize over it in
$\widetilde{O}(n^2)$ time. However, the number of cut queries incurred
this way is quadratic and not near-linear, as promised
in~\Cref{thm:unweighted-cuts}.

However, we can use ideas from \cite{RSW18-value-oracle-mincuts} to do
better. Suppose $G'$ is a contracted graph (i.e., where some nodes in
$G$ have been contracted into each other), and $G'(p)$ is the
``edge-percolation'' graph obtained by sub-sampling each edge of $G'$
independently with probability $p$. Distilling the results of
\cite{RSW18-value-oracle-mincuts} shows that if, for any contraction
$G'$ of $G$, we can obtain a sample from $G'(p)$ using
$\alpha \cdot p\cdot |E(G')|$ queries, then we can compute the min-cut of $G$
using $\alpha \cdot \widetilde{O}(n)$ queries and in $\tO(n^2)$ expected runtime. (A proof appears in 
\Cref{sec:cut-sparsifier-details}.)

To use this result, we show how to sample percolations of contractions
with $\alpha = \widetilde{O}(1)$. 
\begin{restatable}[Contracted Graphs]{lemma}{Contracted}
	\label{lem:contracted-graphs}
	When $n\geq 4$, for any contraction $G'=(U', E')$ of $G$, with its vertex set $U'$
	represented as a partition of the original vertex set $U$, we can sample a subgraph $G'(p)$ where each edge in $G'$ is sampled
	independently with probability $p$ using $|E'|\cdot O(p\log n)+\widetilde{O}(n)$
	queries and $\tO(n^2)$ time in expectation. 
	
	The sampling can be done more generally for any induced subgraph $G''=G'[S]$ for some $S\subseteq U'$ with $|E'(S)|\cdot O(p\log n)+\widetilde{O}(n)$
	queries and $\tO(n^2)$ time.
\end{restatable}
\begin{proof}
	For any regular vertex $u$, if $V_u$ is the super vertex that
	contains it, we will show how to sample edges from
	$\partial(u, U-V_u)$ with each edge sampled independently with
        bias $q\in [0,1]$. Call this random subgraph $\partial_{G'}(u,
        q)$. Given the primitive to sample $\partial_{G'}(u, q)$,
        taking the union $\cup_{u\in U}\partial_{G'}(u, q)$ of such
        samples for each vertex $u\in U$ with $q=1-\sqrt{1-p}$ (so
        that $1-(1-q)^2 = p$) gives $G'(p)$. Because, the probability of any edge $e\in E'$ to
	be selected is $1-(1-q)^2=p$ and the independence is trivial. 
	
	To sample $\partial_{G'}(u, q)$, let $T$ be a random subset of $U-V_u$ with each element sampled independently with probability $q$. Reveal all the edges of $u$ going into $T$ using \Cref{lem:edge-extraction}. This takes $|\partial(u,T)|\cdot O(\log n)$ queries which is $q\cdot |\partial(u, U-V_u )|\cdot O(\log n)$ in expectation and $\tO(n)$ time. Summing over all $u\in U$ gives a total of $|E'|\cdot O(p\log n)+\widetilde{O}(n)$
	queries and $\tO(n^2)$ time in expectation. An edge $\{u,v\}$ for $v\in U\backslash V_u$ is sampled iff $v\in T$. This implies that each edge is sampled independently with probability $q$. Simply replacing $U$ with $S$ gives the extension to induced subgraphs. 
\end{proof}

Combining \Cref{lem:contracted-graphs} with the abstraction of \cite{RSW18-value-oracle-mincuts} discussed above completes our algorithm for min-cuts (\Cref{thm:unweighted-cuts}).

\subsection{Proof of the Structure Primitives Theorem}
\label{sec:proof-struct}

Finally, we turn to the proof of the Structure Primitives result in
\Cref{lem:cut-query-structural}. We will provide a proof sketch here
for the query complexity bounds; see \Cref{app:cut-proofs} for the full proof including
the running time bounds.

\begin{proof}[Proof sketch of \Cref{lem:cut-query-structural}]
	Recall that we want to identify the edges coming out of $u$. The key
	observation is that, for any set $S$ not containing $u$, the sign of
	$\partial(S+u) - \partial(S)$ tells us whether less, equal to, or
	more than half of the edges incident to $u$ go into $S$ (if the
	sign is positive, zero, or negative).
	
	Suppose we find a set $S$ such that \emph{just below} half of the
	edges from $u$ go into $S$ (so that expression has a positive
	sign)---such an $S$ is called a \emph{median set for $u$}. Now if we
	add $v$ to the median set $S$, and the sign of
	$\partial(S+v+u) - \partial(S+v)$ changes, we know that $v$ is a
	neighbor of $u$. In summary, once we have a median set $S$, we can
	find all the neighbors of $u$ not in $S$.
	
	How do we find a median set for $u$? Order all the vertices other
	than $u$ (say this is $v_1, v_2, \ldots, v_{n-1}$, define $S_i$ to
	be the first $i$ elements. If $u$ has at least one edge incident to
	it, the query $\partial(S_0+u) - \partial(S_0)$ has a positive sign,
	whereas $\partial(S_{n-1}+u) - \partial(S_{n-1})$ has a negative
	sign (note that we use \( S_0 = \emptyset \) and \( S_{n-1} + u = U \) here for the sake of exposition, but such trivial cuts are not allowed in the actual comparison model. The full proof in \Cref{app:cut-proofs} handles this correctly and with full rigor). So we can perform binary search to find a point where the sign
	changes, which can be used to find a median set in $\log_2 n$
	comparisons. (In fact, we can find two disjoint median sets---a
	prefix and a suffix of this ordering---and use the two to find all
	the neighbors in some $k$-sized set with $k + O(\log n)$ queries.)
	
	For part~\ref{item:struct2}, the argument builds on the use of such a median set $S$. Given an ordered set $T = \{v_1, \dots, v_r\}$ where $v_j$ is the first neighbor of $u$ in the ordering, we first compute a median set $S$ that is guaranteed to exclude the prefix $\{v_1, \dots, v_j\}$. Then, we binary search over the chain 
	\[
	S \subseteq S \cup \{v_1\} \subseteq \cdots \subseteq S \cup (T \setminus S)
	\]
	to locate the point at which the marginal changes sign. This transition occurs precisely at the set $S \cup \{v_1, \dots, v_{j-1}\}$, thus identifying $v_j$ as the first neighbor of $u$ in $T$.
\end{proof}

\subsection{Sparsifiers for Heavy Subgraph Problems}
\label{sec:sparsifier-gen}

The work of \cite{EHW15-uniformsampling} considers a broad class of
graph optimization problems, which they call \emph{heavy subgraph}
problems; these include densest subgraph, densest bipartite subgraph,
and $d$-max cut. For these problems, they show that uniformly sampling
$\widetilde{O}(n / \eps^2)$ edges without replacement from the
underlying graph $G$ produces a subgraph $H$ such that running any
$\alpha$-approximation algorithm for the problem on $H$ yields (after
appropriate rescaling) an $(\alpha - \eps)$-approximate solution for
the original graph, with high probability. In other words, the sampled
subgraph preserves the structure of the problem up to a small loss in
accuracy, and approximate solutions on the sparsifier $H$ translate to
approximate solutions on the full graph $G$. We show how to sample a fixed number of edges from $G$ uniformly using the following lemma: 
\begin{restatable}[Sampling Lemma]{lemma}{Sampling}
	\label{lem:degree-estimation}
	When $n\geq 4$, given cut comparison queries, 
	for any $k=\widetilde{\Omega}(1)$, we can
	sample $k$ edges uniformly without replacement using
	$\widetilde{O}(n+k)$ queries and $\widetilde{O}(n^2)$ time with high probability. 
\end{restatable}
\begin{proof}
	Consider the process that at time step $t$ (starting from $t=0$)
	samples $G(p_t)$ for $p_t:= 2^t/n^2$ until the sample $G(p_t)$ at
	$t=r$ has at least $k$ edges. Then, subsample $k$ edges uniformly
	from the sample $G(p_r)$.
	
	Because edges are sampled i.i.d.\ from the Bernoulli
	$\text{Ber}(p_t)$ distribution at each step $t$, the symmetry
	implies that the subsample of edges is a uniform sample of $k$
	edges. Using \Cref{lem:contracted-graphs}, each sample $G(p_t)$
	takes $p_t \cdot \widetilde{O}(m)+\widetilde{O}(n)$ queries and
	$\tO(n^2)$ time in expectation. Summing this over $0\leq t\leq r$
	gives an upper bound of
	$2p_r\cdot \widetilde{O}(m)+\widetilde{O}(nr)$ queries and
	$\tO(n^2r)$ time. Using the fact that the probability that $G(p)$
	has less than $k$ edges for $p=2k/m$, is at most $e^{-k/4}$, we can
	conclude that $p_r\leq 4k/m$ with probability at least
	$1-e^{-k/4}$. So the expected number of queries is upper bounded by
	$\widetilde{O}(nr+k)=\widetilde{O}(n+k)$ and the running time by
	$\tO(n^2r)=\tO(n^2)$ using the fact that $r\leq O(\log n)$.
\end{proof}
Using \Cref{lem:degree-estimation} with $k=\widetilde{O}(n / \eps^2)$,
we can obtain a random subgraph $H$ of the unknown graph $G$ to get
such a sample of edges using $\widetilde{O}(n/\eps^2)$ queries, and
hence apply existing approximations for all heavy subgraph problems.


The results in this section do not generalize to the weighted case when the edges of the graph have unknown non-negative weights. We address this issue by giving some preliminary results under additional structural assumptions and slightly stronger oracle models—see \Cref{sec:weighted-cuts}—but the general question remains wide open.

See \Cref{sec:useful-examples} for some illustrative examples which show some separation between the capabilities of the cut comparison and cut value query oracles.


\bigskip
\noindent {\bf \LARGE Appendix}

\appendix
\section{Omitted Content from \Cref{sec:gen_set_system}}
\label{sec:appendix-geometric}



In this section, we give omitted proofs and other discussion
from~\Cref{sec:gen_set_system}. In particular, we show that the
``certification'' problem, where we have to write the shortest
certificate of some point $y^*$ being the optimizer of $\ip{w^*,x}$
over points $x \in \cP$, can be directly related to the conic
dimension. We present this discussion since it helps motivate and
demystify the definition further. We then give the proofs omitted from \Cref{sec:gen_set_system}.

The notion of conic dimension is intimately related the notion
of  
\emph{inference dimension} given in
\cite{KLMZ17-comp-based-classification}. They defined it as being the
size of a largest ``inference-independent'' set where none of the
inequalities are implied by the others. Our definition is very similar
to theirs, with two minor differences: (a)~we restrict to linear
optimization, and (b)~we consider sequences where each comparison is
not implied by the previous ones. Since implications using linear
inequalities has a natural geometric visualization as a cone, we call
it \emph{\CDfull}.

\subsection{The Certification Problem}
\label{sec:certification}
Before addressing the optimization problem, we ask whether it is even
possible to certify optimality efficiently. Suppose we wish to prove
that some point $y\in \cP$ is the optimal solution, i.e., 
$y\in \arg\min_{x \in \cP} \langle w^*, x \rangle$, or equivalently, that
\begin{align}
	\ip{w^*, x-y} \geq 0 \quad \forall x\in \cP.
\end{align}
We want to use as few comparisons as possible. 

\begin{enumerate}
	\item A $y$-\emph{certificate} is a collection $\cC$ of index pairs
	$(i,j)$ such that for all $w \in \mathbb{R}^d$, we have that
	\begin{align}
		\left(\langle w, x_i - x_j\rangle \geq 0 \quad \forall (i,j)\in \cC\right) \Rightarrow \left(\langle w, x - y\rangle \geq 0 \quad \forall x \in \cP\right) 
	\end{align}
	\item A certificate $\cC$ is \emph{valid} if
	$\langle w^*, x_i - x_j \rangle \geq 0$ for all $(i,j) \in \cC$.
\end{enumerate}
Hence a valid $y$-certificate implies that $y$ is a minimizer of
$\ip{ w^*, x}$ among points in $\cP$. In fact, the minimum number of
queries needed to find an optimal solution is lower bounded by the
minimum size of a valid certificate.

Recall that given a set
$V \sse \RR^d$ of vectors, the cone generated by $V$ is the set of all
vectors that can be written as positive linear combinations of vectors
from $V$; this is denoted by $\cone(V)$. Formally,
\[ \cone(V) = \left\{x \in \mathbb{R}^d : x = \sum_{v \in V} \alpha_v
\, v, \; \alpha_v \geq 0\right\}. \]
The following lemma gives a characterization of $y$-certificates
based only on the geometry of the point set $\cP$. 
\begin{lemma}
	\label{lem:y-certif-cone-eqiv}
	A set $\cC \subseteq [N] \times [N]$ is a $y$-certificate iff $\cP \subseteq y + \cone\left(\{x_i - x_j\}_{(i,j)\in \cC}\right)$.
\end{lemma}
\begin{proof}
	Suppose $\cC$ is a $y$-certificate. Then, the definition implies that the system:
	\begin{align}
		\label{eqn:certificate-equiv}
		&\langle w, x_i - x_j \rangle \geq 0 \quad \forall (i,j) \in \cC \\
		&\langle w, x - y \rangle < 0 \nonumber
	\end{align}
	is infeasible for any $x\in \cP$. By Farkas' lemma, this implies
	that $x - y \in \cone\left(\{x_i - x_j\}_{(i,j)\in \cC}\right)$.
	
	Conversely, if $x - y \in \cone\left(\{x_i -
	x_j\}_{(i,j)\in \cC}\right)$ and $\langle w, x_i -
	x_j\rangle \geq 0$ for all $(i,j)\in \cC$, then $\langle w,
	x - y \rangle \geq 0$ follows immediately. \qedhere
\end{proof}

\begin{defn}
	\label{defn:induced-certificate}
	For a sequence $\sigma=(x_{i_1},x_{i_2},\dots, x_{i_k})$, the \emph{induced certificate} is defined as $\cC_{\sigma}=\{(i_{\ell},i_{\ell+1}): 1\leq \ell\leq k-1\}$  
\end{defn}

\subsubsection{Basic Subsequences}
\label{sec:basic-subsequences}

For a set $\cY\subseteq \RR^d$ with $\CDmath(\cY)=k$ and any sequence $\sigma = (y_1, \dots, y_n) \in \cY^n$, consider a
process that maintains a subsequence $\pi_t$ of $\sigma$ for each
$0 \leq t \leq n$. We start with the empty sequence $\pi_0 = \ip{}$,
and at step $t+1$, we update
\[
\pi_{t+1} = 
\begin{cases}
	\pi_t \circ y_{t+1}, & \text{if } y_{t+1} - y_1 \notin \envelope(\pi_t) \\
	\pi_t, & \text{otherwise}.
\end{cases}
\]
An analogy can be made to Kruskal's algorithm, where we add the next
element in the sequence precisely when it is not conically spanned by
the envelope of the current set. We observe:
\begin{enumerate}[nosep]
	\item the sequence $\pi_t$ is a prefix of $\pi_{t+1}$, and moreover,
	$\pi_{t+1}$ contains at most one more element.
	\item $|\pi_n| \leq k$.
\end{enumerate}
The first observation is immediate from the construction, and the
second follows by contradiction: if $|\pi_n| > k$, then some
subsequence of size $k+1$ contradicts the definition of \CDfull. Let $B(\sigma) := \pi_n$ be the subsequence of $\sigma$
obtained by this process; we call this the ``basis'' of $\sigma$. 

\begin{lemma}
	\label{lem:basis-cone-property}
	Given a sequence $\sigma = (y_1, \dots, y_n) \in \cY^n$, we have:
	\begin{align}
		\cone(\{y_j - y_1\}_{1 \leq j \leq n}) \subseteq \envelope(B(\sigma)) \subseteq \envelope(\sigma).
	\end{align}
\end{lemma}

\begin{proof}
	For the rightmost inclusion, suppose $B(\sigma) = (y_{i_1}, \dots, y_{i_k})$. Then,
	\begin{align}
		\envelope(B(\sigma)) = \cone(\{y_{i_{\ell+1}} - y_{i_\ell}\}_{1 \leq
			\ell \leq k-1}) \subseteq \cone(\{y_j - y_i\}_{1 \leq i \leq j \leq
			n}) = \envelope(\sigma).
	\end{align}
	For the leftmost inclusion, let $\pi_t := B(\sigma^{1:t})$ denote the basis of the prefix at time $t$. Consider any $y_t \in \sigma$:
	\begin{itemize}
		\item If $y_t \notin \pi_n = B(\sigma)$, then it was excluded by the base construction algorithm, meaning:
		\[
		y_t - y_1 \in \envelope(\pi_{t-1}) \subseteq \envelope(\pi_n) = \envelope(B(\sigma)).
		\]
		\item If $y_t \in \pi_n$, then it is one of the elements of the basis, and we can write:
		\[
		y_t - y_1 =\sum_{\ell: i_{\ell}\leq t}(y_{i_{\ell+1}}-y_{i_\ell}) \in \cone(\{y_{i_{\ell+1}} - y_{i_\ell}\}_{1 \leq \ell \leq k-1}) = \envelope(B(\sigma)).
		\]
	\end{itemize}
	
	Hence, in both cases, $y_t - y_1 \in \envelope(B(\sigma))$, completing the proof.
\end{proof}

We can now relate the size of valid $x^*$-certificates to the \CDfull of the underlying point set.
\begin{lemma}
	\label{lem:size-to-dimension}
	If some point set $\cY \subseteq \mathbb{R}^d$ has \CDfull $k$, and $y^* = \arg\min_{y \in \cY} \ip{w^*,y}$, then there
	exists a valid $y^*$-certificate $\cC$ of length $|\cC| \leq k - 1$.
\end{lemma}
\begin{proof}
	Order the points in $\cY$ as the sequence
	$\sigma = \langle y^* = y_1, y_2, \ldots, y_n \rangle$ such that
	$\langle w^*, y_i \rangle \leq \langle w^*, y_{i+1} \rangle$ for all
	$i$, and let $B(\sigma)=(y_{i_1}, \dots, y_{i_k})$ be the basic subsequence
	corresponding to $\sigma$. We
	claim that the induced certificate $\cC:=\cC_{B(\sigma)}$ is a valid $y^*$-certificate. Notice that $\cC$ is defined such that 
	\[
	\envelope(B(\sigma))=\cone(\{y_i - y_j\}_{(i,j)\in \cC}).
	\]
	Using \Cref{lem:basis-cone-property}, we have  $\cY\sse y_1+\cone(\{y_i - y_j\}_{(i,j)\in \cC})$,
	which means that $\cC$ is a valid certificate using
	\Cref{lem:y-certif-cone-eqiv}. Moreover, by construction,
	$(i,j) \in C$ means $\ip{w^*, y_j - y_i} \geq 0$, which in turn
	means that the certificate is valid.
\end{proof}


\subsubsection{Proof of \Cref{lem:alg-progress}}
\label{sec:proof-alg-progress}

\Sieve*

\begin{proof}[Proof of~\Cref{lem:alg-progress}]
	For the analysis, consider a sequence $(x_1, \dots, x_N)$ be such
	that $\langle w^*, x_i \rangle \leq \langle w^*, x_{i+1} \rangle$
	for all $1\leq i\leq N-1$. Note that $\sigma$ is distributed as a
	random subsequence of $(x_1, \dots, x_N)$, where each element is
	selected independently with probability $2k/N$.
	
	Consider a slightly modified, more lenient version of the algorithm
	that eliminates a point $x_t \in \cP \setminus \{y\}$ in
	Step~\ref{item:alg-step-2} only if
	$x_t - y \in \envelope(B(\sigma_{t-1}))$ where $\sigma_t$ is the
	prefix of $\sigma$ restricted to elements $\{x_i\}_{1\leq i\leq t}$;
	recall the definition of the basic subsequence $B(\cdot)$ from
	\Cref{sec:basic-subsequences}. This modification ensures that
	elimination of $x_t$ depends only on the coin tosses at indices in
	$[t-1]$ and the sequence $(x_1,\dots,x_t)$. This modified algorithm
	eliminates no more points than the original algorithm because
	$\envelope(B(\sigma_{t-1}))\sse \envelope(\sigma_{t-1})\sse
	\envelope(\sigma)$---each inequality just uses that the cone of a
	smaller set of vertices is smaller. Hence, it suffices to prove that
	the expected number of points eliminated by the modified algorithm
	is at least $N/2$.
	
	We now use the principle of deferred randomness to analyze this
	version of the algorithm. Let $E_t$ be the ``evolving set'' of elements in $\{x_i\}_{1\leq i\leq t}$ that are not eliminated by the modified algorithm after it has seen the elements $x_1,x_2\dots,x_t$ in that order. Remember that $\sigma_t$ is the subsequence of $(x_1,\dots,x_t)$ corresponding to elements that are sampled. We can simulate the process of generating the sets $E_t$ and $\sigma_t$ in the following way:
	\begin{enumerate}
		\item Start with $E_0, \sigma_0=\ip{}$.
		\item At step $1\leq t\leq N$, toss a biased coin with success probability $2k/N$. Update 
		\begin{align*}
			\sigma_{t} &= 
			\begin{cases}
				\sigma_{t-1} \circ x_{t}, & \text{if coin flip succeeds at } $t$ \\
				\sigma_{t-1}, & \text{otherwise}.
			\end{cases} 
			\intertext{and}
			E_{t} &= 
			\begin{cases}
				E_{t-1} \circ x_{t}, & \text{if } x_{t} - y \notin \envelope(B(\sigma_{t-1})) \\
				E_{t-1}, & \text{otherwise}.
			\end{cases}
		\end{align*}
		where $y$ is the first element sampled.
	\end{enumerate}
	Observing that 
	\begin{align*}
		|B(\sigma_{t})|-|B(\sigma_{t-1})| &=\mathbbm{1}[x_t-y \notin \envelope(B(\sigma_{t-1})) \land \textnormal{coin flip succeds at }t] \\
		&=(|E_t|-|E_{t-1}|)\cdot\mathbbm{1}[\textnormal{coin flip succeds at }t].\\
		\intertext{
			Taking expectations on both sides gives 
		}
		\EE[|E_t|-|E_{t-1}|]&=\frac{N}{2k}\cdot \EE[|B(\sigma_{t})|-|B(\sigma_{t-1})|].
		\intertext{Summing this over $1\leq t\leq N$ gives}
		\EE[|E_{N}|]&=\frac{N}{2k}\cdot \EE[|B(\sigma)|]\leq
		\frac{N}{2k}\cdot k =\frac{N}{2}.  \qedhere
	\end{align*}
\end{proof}

\subsection{The \CDfull of the Boolean Point Sets}
\label{sec:boolean-points}

An important special case is that of point sets consisting of Boolean
vectors in $\{0,1\}^d$, i.e., of subsets of the hypercube. For
completeness, we present a repackaging of the bound on the \CDshort from
\cite{KLMZ17-comp-based-classification}. To begin, we recall the fact that all subset sums of any linearly independent set of vectors are distinct, and extend this fact to the case of conic independence. The bound on the conic dimension of Boolean point sets then follows by a counting argument.

\begin{lemma}
\label{lem:conic-indep-subsetsum}
If $\sigma = (y_1, y_2, \dots, y_{k+1})$ is a conically independent sequence, then all subset sums 
\[
\sum_{i \in S} (y_{i+1} - y_i), \quad S \subseteq [k],
\]
are distinct.
\end{lemma}

\begin{proof}
Suppose for contradiction that the claim fails, and let $k$ be the smallest integer for which a counterexample exists. 
The statement is trivial for $k = 1$, so $k \ge 2$.  
Let $\sigma = (y_1, \dots, y_{k+1})$ be such a minimal counterexample.  
Then there exist distinct subsets $S, T \subseteq [k]$ satisfying
\[
\sum_{i \in S} (y_{i+1} - y_i) = \sum_{i \in T} (y_{i+1} - y_i).
\]
By minimality, the symmetric difference $S \triangle T$ must contain $k$; otherwise, restricting to the first $k-1$ elements would yield a smaller counterexample.  
Without loss of generality, assume $k \in S$ and define $S' = S \setminus \{k\}$.

Rearranging gives
\[
y_{k+1} - y_k = \sum_{i \in [k-1]} \big( \mathbbm{1}_{S'}(i) - \mathbbm{1}_T(i) \big) (y_{i+1} - y_i).
\]
Substituting into the telescoping sum 
\[
y_{k+1} - y_1 = (y_{k+1} - y_k) + \sum_{i \in [k-1]} (y_{i+1} - y_i),
\]
we obtain
\[
y_{k+1} - y_1 = \sum_{i \in [k-1]} \big( 1 + \mathbbm{1}_{S'}(i) - \mathbbm{1}_T(i) \big) (y_{i+1} - y_i),
\]
so $y_{k+1} - y_1$ lies in the conic hull of $\{y_2 - y_1, \dots, y_k - y_{k-1}\}$, contradicting the conic independence of $\sigma$.
\end{proof}

\begin{lemma}[\cite{KLMZ17-comp-based-classification}]
\label{lem:conic-dim-boolean-upperbound}
The conic dimension of any subset of $\{0,1\}^d$ is at most $O(d \log d)$.  
In fact, it is the smallest $k$ such that $2^k > (2k+1)^d$.
\end{lemma}

\begin{proof}
Let $\sigma = (y_1, \dots, y_{k+1})$ be a conically independent sequence of Boolean points.  
By \Cref{lem:conic-indep-subsetsum}, the $2^k$ subset sums 
\[
\sum_{i \in S} (y_{i+1} - y_i), \quad S \subseteq [k],
\]
are all distinct.  
Each such sum is an integer vector in $[-k, k]^d$, of which there are at most $(2k+1)^d$ possibilities.  
Hence $2^k \le (2k+1)^d$, implying that $k = O(d \log d)$.
\end{proof}

\subsection{Beyond the Hypercube}
\label{sec:beyond-hypercube}
In this section, we consider extensions of the above claim for the Boolean hypercube: to the setting of hypergrids, and to getting $\eps$-approximate solutions for other bounded sets. 

\paragraph{Hypergrids and Bounded-Precision Solutions}

A more general version of \Cref{lem:conic-dim-boolean-upperbound} when points belong to $\{0,1,\dots, N-1\}^d$ shows that the 
\CDfull is shown to be at most $O(d\log (Nd))$. The proof is  very similar to that of \Cref{lem:conic-dim-boolean-upperbound} above;
see \cite[Lemma 4.2]{KLMZ17-comp-based-classification}. 

\paragraph{$\eps$-Approximate Solutions for Bounded Sets}
For bounded sets $\cP \subset \BB_d = \{x \in \RR^d; \norm{x}_2\leq 1\}$ it is possible to bound an approximate version of the \CDfull, which leads to an algorithm that returns an $\eps$-approximate point. Given a comparison oracle, the algorithm returns $\widehat x \in \cP$ such that $\langle w^*, \widehat x \rangle \leq \langle w^*, x\rangle + \eps$ for all $x \in \cP$.

For the following definition, we will use the notion of the distance: given a set $S \subset \RR^d$ and a point $x \in \RR^d$, define $\dist(x,S) = \min_{y \in S} \norm{x-y}_2$.

\begin{defn}
	\label{defn:approx-conic-dimension}
	The \emph{$\eps$-approximate \CDfull} of a \emph{set}
	$\cY \subseteq \mathbb{R}^d$ denoted by $\CDmath_\eps(\cY)$ is the largest integer $k$ for which
	there exists a length-$k$ sequence
	$\sigma = (y_1, \dots, y_{k}) \in (\cY)^k$ of points from $\cY$, such
	that for each $t \in \{2, \ldots, k\}$ we have
	\begin{align}
		\dist(y_t - y_1, \envelope(\sigma^{1:t-1})) \geq \eps.
	\end{align}
\end{defn}

All the notions previously defined extend to the approximate \CDshort with the natural changes. 
We can extend the notion of a basic subsequence $B_\eps(\sigma)$ following the same Kruskal-like procedure: we add an element to the basic subsequence if its distance to the previously spanned elements is at least $\eps$. 

If $k = \CDmath_\eps(\cY)$ we can apply Algorithm \ref{alg:sample-remove-recurse} with two modifications: after the first iteration, always include the minimum point from the previous iteration in the sample; in step 2 eliminate all points $x\in \cP \setminus \{y\}$ such that $\dist(x-y, \envelope(\sigma)) < \eps$ obtaining the following results:

\begin{cor}
	If $\norm{w^*} \leq 1$ and $k = \CDmath_\eps(\cY)$, the the variant of Algorithm \ref{alg:sample-remove-recurse} described above returns a point $\hat x$ such that $\langle w^*, \hat x \rangle \leq \langle w^*,  x \rangle + \eps$ for all $x \in \cP$ using $O(k \log k \log |\cP|)$ comparisons.
\end{cor}

\begin{proof} Observe that if $y$ is the smallest point in the sampled subsequence then if $\dist(x-y, \envelope(\sigma)) < \eps$ then let $z \in y + \envelope(\sigma)$ be a point such that $\norm{x-z}_2 \leq \eps$. We know that $$\langle w^*,y \rangle \leq \langle w^*,z \rangle = \langle w^*,x \rangle + \langle w^*,z-x \rangle \leq \langle w^*,x \rangle + \eps$$
	Hence we only eliminate points that are at no better than the point $y$ by more than $\eps$. The modification in step 1 guarantees that the points $y$ selected in each iteration as increasingly better. The remainder of the proof works without any change.
\end{proof}

Finally we bound the approximate \CDshort:

\begin{lemma}
	\label{lem:approx-conic-dim-upperbound}
	The $\eps$-approximate \CDfull of any set of points $\cP \subset \BB_d$ is the smallest $k$ such that
	$2^k (\eps/2)^d > (2k+1)^d$.
\end{lemma}
\begin{proof}
	The proof follows by replacing the pigeonhole argument in Lemma \ref{lem:conic-dim-boolean-upperbound} by a volumetric argument. We again consider the set:
	\begin{align}
		\Big\{\sum_{i \in [k]} \beta_i (y_{i+1} - y_i) : \beta_i \in \{0,1\}\Big\} \subseteq 2k \cdot \BB_d
	\end{align}
	i.e, the ball of radius $2k$ in $\RR^d$, which has volume $(2k)^d \cdot \text{Vol}(\BB_d)$. Now, for each vector $\beta \in \{0,1\}^k$, consider the ball of radius $\eps/2$ around $\sum_{i \in [k]} \beta_i (y_{i+1} - y_i)$, which have total volume $2^k (\eps/2)^d \cdot \text{Vol}(\BB_d)$. Each of those balls is contained in the ball of radius $2k+1$ around zero. If $2^k (\eps/2)^d > (2k+1)^d$ then two of the smaller balls must intersect, so there exist vectors $\beta, \gamma \in \{0,1\}^k$ such that:
	\begin{align}
		\sum_{i \in [k]} (\beta_i - \gamma_i)(y_{i+1} - y_i) =z,
	\end{align}
	with $\norm{z}_2 < \epsilon$. Let $t_0$ be the largest index where
	$\beta_{t_0} \neq \gamma_{t_0}$, and assume $\beta_{t_0} = 0$,
	$\gamma_{t_0} = 1$. Doing the same manipulations as in Lemma \ref{lem:conic-dim-boolean-upperbound} we obtain:
	\begin{align}
		z + y_{t_0+1} - y_1 = \sum_{i \in [t_0-1]} (\beta_i - \gamma_i + 1)(y_{i+1} - y_i)
	\end{align}
	Since $\beta_i, \gamma_i \in \{0,1\}$, the coefficients are
	non-negative, so
	$$\dist(y_{t_0+1} - y_1 , \envelope(y_1, \dots, y_{t_0})) \leq \norm{z}_2 < \eps$$
	which shows that that $\eps$-approximate \CDfull is at most $k$.
\end{proof}


\section{Omitted proofs from \Cref{sec:ellipsoid}}
\label{sec: ellipsoid-omitted-proofs}

\begin{proof}[Proof of \Cref{lem:arrangement-cells}]
	For the first part, observe that $C_\pi$ is the intersection of $N - 1$ closed half-spaces, and is thus a convex polyhedron.
	
	Assume $C_\pi$ is full-dimensional. Then its interior, $\operatorname{int}(C_\pi)$, is non-empty. For any $w \in \operatorname{int}(C_\pi)$, strict inequalities $\langle w, x_{\pi(1)}\rangle < \langle w, x_{\pi(2)}\rangle < \dots < \langle w, x_{\pi(N)}\rangle$ hold. Consequently, $\operatorname{int}(C_\pi)$ does not intersect any hyperplane in $\mathcal{A}$. Since $C_\pi$ is convex, it is connected.
	To show $C_\pi$ is a cell, we check maximality. Suppose $S \subseteq \mathbb{R}^d$ is a connected region such that $C_\pi \subseteq S$ and $\operatorname{int}(S) \cap \mathcal{A} = \emptyset$. Since $S$ is connected and its interior avoids all hyperplanes $h_{\{i,j\}}$, the sign of $\langle w, x_i - x_j \rangle$ must be constant for all $w \in \operatorname{int}(S)$. Because $\operatorname{int}(C_\pi) \subset \operatorname{int}(S)$, this sign pattern is determined by $\pi$. Thus, every $w \in S$ must satisfy the constraints defining $C_\pi$, implying $S \subseteq C_\pi$. Therefore, $C_\pi$ is a maximal connected region whose interior avoids $\mathcal{A}$, making it a cell.
	
	Conversely, let $C$ be any cell of the arrangement $\mathcal{A}$. Since cells are full-dimensional, $\operatorname{int}(C)$ is non-empty. Pick any $w \in \operatorname{int}(C)$; the values $\{\langle w, x_k \rangle\}_{k=1}^N$ are distinct and induce a permutation $\pi \in S_N$. Since $C$ is connected and $\operatorname{int}(C) \cap \mathcal{A} = \emptyset$, the ordering $\pi$ is constant over $\operatorname{int}(C)$. Thus $C \subseteq C_\pi$. Since $C$ is a maximal region with this property and $C_\pi$ is a region satisfying the property (as shown above), it must be that $C = C_\pi$.
	
	For the second part regarding partial orders, $C_\sigma$ is the intersection of the half-spaces $H_{j,i}$ for all pairs $i \preceq_\sigma j$, which implies it is a convex polyhedron. To prove the decomposition, consider any $w \in C_\sigma$. The scalar values $\{\langle w, x_k \rangle\}_{k=1}^N$ induce a total ordering. Since $w$ satisfies the constraints of $\sigma$, this total ordering is a linear extension of $\sigma$. Thus, there exists some $\pi \in \mathcal{L}(\sigma)$ such that $w \in C_\pi$. Conversely, if $w \in C_\pi$ for some $\pi \in \mathcal{L}(\sigma)$, then by transitivity, $w$ satisfies $\langle w, x_i \rangle \leq \langle w, x_j \rangle$ for all $i \preceq_\sigma j$, implying $w \in C_\sigma$. Therefore, $C_\sigma = \bigcup_{\pi \in \mathcal{L}(\sigma)} C_\pi$.
\end{proof}

\begin{proof}[Proof of \Cref{lem:sigma^*}]
	By definition, the partial order $\sigma^*$ contains exactly those relations $i \preceq_{\sigma^*} j$ for which $\langle w^*, x_j - x_i \rangle > 0$. Since $C_{\sigma^*}$ is defined by the intersection of half-spaces $\langle w, x_j - x_i \rangle \geq 0$ for all $i \preceq_{\sigma^*} j$, and $w^*$ satisfies all these inequalities strictly, $w^*$ must lie in the interior of $C_{\sigma^*}$.
	
	For the second part, we prove the contrapositive: $\langle w^*, x_i-x_j \rangle > 0 \implies \langle \widehat{w}, x_i-x_j \rangle > 0$.
	Suppose $\langle w^*, x_i - x_j \rangle > 0$. By the definition of $\sigma^*$, this strict inequality implies the relation $j \preceq_{\sigma^*} i$. Consequently, every point $w \in C_{\sigma^*}$ must satisfy $\langle w, x_i - x_j \rangle \geq 0$. Since $\widehat{w}$ is in the interior of $C_{\sigma^*}$, it must strictly satisfy the inequalities implied by the partial order. Therefore, $\langle \widehat{w}, x_i - x_j \rangle > 0$.
\end{proof}

\begin{proof}[Proof of \Cref{cor:CCO-compatible}]
	By the definition of the CCO output for $\widehat{w}$, the sequence is non-decreasing with respect to $\widehat{w}$, meaning $\langle \widehat{w}, y_i - y_{i+1} \rangle \leq 0$ for all $1 \leq i \leq k-1$.
	Applying \Cref{lem:sigma^*}, the implication
	\[ \langle \widehat{w}, x_u - x_v \rangle \leq 0 \implies \langle w^*, x_u - x_v \rangle \leq 0 \]
	holds for all pairs from $\cP$. Therefore, we have $\langle w^*, y_i - y_{i+1} \rangle \leq 0$ for all $i$. By transitivity, $\langle w^*, y_i - y_j \rangle \leq 0$ for all $1 \leq i \leq j \leq k$.
	
	Finally, using the property that $\cP \subseteq y_1 + \envelope(y_1, \dots, y_k)$, and observing that $w^*$ makes a non-negative inner product with the generators of this envelope (as implied by the sorted order), it follows that $y_1 \in \operatorname*{argmin}_{x\in \cP}\langle w^*, x \rangle$.
\end{proof}

\section{Omitted proofs from \Cref{sec:GSL}}
\label{sec:GSL-appendix}

\begin{proof}[Proof of \Cref{lem:scalar-encoding}]
  The upper bound on the value is immediate, and the non-negativity follows since
  \[
    \sum_{j=1}^s (3nM)^{j-1} w_j \ge -M \sum_{j=1}^s (3nM)^{j-1} > -(3nM)^s.
  \]
  For any \(T \subseteq U\),
  \[
    \sum_{i \in T} g(w_i)
    = (3nM)^s \cdot |T| + \sum_{j=1}^s (3nM)^{j-1} \cdot \left(\sum_{i \in T} w_{i,j}\right).
  \]
  Here $|T| \in [0,n]$ and each
  $\sum_{i \in T} w_{i,j} \in [-nM, nM]$.  Thus,
  $\sum_{i \in T} g(w_i)$ is a base-$(3nM)$ encoding of the
  tuple
  $(|T|, \sum_{i \in T} w_{i,1}, \ldots, \sum_{i \in T} w_{i,s})$,
  which can be uniquely decoded. Hence, equality of the encoded sums
  holds iff the corresponding vector sums are equal.
\end{proof}


\LinSep*

\begin{proof}[Proof of \Cref{lem:dif-weight-basis}]
  Consider the polynomial $P(x)$ defined in~(\ref{eq:3}):
  \[
    P(x) = \det\bigg(\sum_{i=1}^n x^{c_i} v_i v_i^T\bigg) = \sum_{S
      \in \cF} x^{\sum_{i \in S} c_i} \det(V_S)^2 = \sum_{S \in \cF}
    x^{c(S)} \det(V_S)^2.
  \]
  As established, the set of exponents with non-zero coefficients in
  $P(x)$ is precisely the set of basis costs, $\{c(S)\mid S\in \cF\}$.

  Assume there exists a basis $S^*$ such that its cost $c = c(S^*)$ is
  not in $Z$. Our goal is to find such a $c$. Instead of explicitly
  computing $P(x)$ which has degree can be $O(nM)$, we compute
  $Q(x) = P(x) \pmod{x^p-1}$ for a carefully chosen prime $p$.
  Indeed, we would like a prime $p$ that does not divide any of the
  integers in the set
  $D = \{z_i - z_j \mid i \neq j \in [\ell]\} \cup \{c - z_j \mid j
  \in [\ell] \}$.  If $p$ satisfies this property, then all residues
  $\{z_i \pmod p\}_{i \in [\ell]}$ are distinct, and $c \pmod p$ is
  distinct from all of them.

  This reduces the original problem to a smaller one: finding an
  exponent $c'$ in $Q(x)$ (which has degree at most $p-1$) such that
  $c' \notin \{z_i \pmod p\}_{i \in [\ell]}$. This new problem can be
  solved in $\poly(n, p)$ time from the warmup proof, which also
  recovers the witness basis $S$ using the self-reductibility approach.

  The final step is to show that a ``small'' prime $p$ with this
  property exists and can be found efficiently. The set $D$ contains
  $O(\ell^2)$ integers. Each integer in $D$ is bounded in magnitude by
  $O(nM)$ (since $c_i, z_j \le M$). The product of all these non-zero
  integers has at most $L := O(\ell^2 \log (nM))$ distinct prime
  factors.
  
  By a simple pigeonhole argument, a prime $p$ that does not divide
  any of these integers must exist among the first $L + 1$ primes. The
  prime number theorem tells us that the $t$-th prime has magnitude
  $O(t \log t)$, the smallest such $p$ has a value of $O(L)$. This
  prime can be found deterministically in $\poly(L)$ time. Hence, the
  overall runtime of the self-reduction algorithm is
  $\poly(L) = \poly(n, \ell, \log M)$, as required. Conversely, if the
  search through all primes in this range yields no such exponent, we
  can conclude that no basis $S$ with $c(S) \notin Z$ exists.
\end{proof}

\section{Omitted proofs from \Cref{sec:minimum-cut}}
\label{app:cut-proofs}

We begin with a simple observation allowing us to test, for a
vertex $u$ and any non-empty set $S$ that is a proper subset of $U-u$, whether a majority of $u$'s
neighbors lie inside $S$. 
\begin{obs} Suppose we perform a cut comparison query on the pair $S$
	and $S+u$. Then the result of the query is:
	\label{obs:query_observation}
	\begin{align}
		\textnormal{sign}(|\partial(S)| - |\partial(S +u)|) =
		\textnormal{sign}\left(|\partial(u, S)| -
		\frac{1}{2}|\partial(u)|\right). \label{eq:1}
	\end{align}
\end{obs}

\begin{proof}
	Indeed, 
	\begin{align*}
		|\partial(S)| - |\partial(S +u)| &= |\partial(u, S)| - |\partial(u, U \setminus (S +u ))| \\
		&= 2|\partial(u, S)| - |\partial(u)|.
	\end{align*}
	Therefore, the sign of the query $(S, S +u )$ indicates whether a
	majority of $u$'s edges go into $S$.
\end{proof}

The above observation allows us to reason about neighborhood structure. We now formalize how to use this to isolate individual neighbors.

\begin{lemma}[tipping point]
	\label{lem:cut_query_technical}
	For $u \in U$, consider a set $\emptyset \neq S \sse U \setminus \{u\}$, and a
	sequence $v_1, v_2, \ldots, v_t$ of vertices in
	$U \setminus (S \cup \{u\})$. Define
	$S_i = S \cup \{v_1, \ldots, v_i\}$. Suppose $S_t\neq U-u$ and 
	the comparisons of the queries $(S_0, S_0 +u)$ and
	$(S_t, S_t +u )$ differ. Then there exists an index $i \in [t]$
	such that
	\begin{align}
		\textnormal{sign}(|\partial(S_{i-1})| - |\partial(S_{i-1} +u )|) <
		\textnormal{sign}(|\partial(S_i)| - |\partial(S_i +u )|). \label{eq:2}
	\end{align}
	The corresponding vertex $v_i$ must be a neighbor of $u$, and can
	be identified using $O(\log t)$ queries and $\tO(n)$ time via binary search.
\end{lemma}

\begin{proof}
	By \Cref{obs:query_observation}, the
	$\textnormal{sign}(|\partial(S)| - |\partial(S +u)|)$ is monotonic
	in $S$, since the quantity $|\partial(u, S)|$ increases as $S$
	grows. Therefore, if the sign changes between $S_0$ and $S_t$, then
	by monotonicity, it must strictly increase at some index $i$, which
	proves (\ref{eq:2}). Moreover, using the equality in (\ref{eq:1})
	again, (\ref{eq:2}) can be rewritten as 
	\begin{align}
		|\partial(u, S_{i-1})| < |\partial(u, S_i)|,
	\end{align}
	and hence $u$ is adjacent to the vertex $v_i$. Finally, a binary
	search over the chain finds such an $i$ using $O(\log t)$
	queries. The binary search performs $O(\log t)$ queries, each
	requiring $O(n)$ time to write bit vectors, for a total of
	$O(n \log t) = \tO(n)$. The additional binary search logic takes
	$O(\log t)$ time.  
\end{proof}

We now apply these ideas to construct balanced sets around each vertex.

\begin{lemma}[median sets]
	\label{lem:medians-u}
	For any vertex $u$ with $\partial(u) \neq \emptyset$, and an ordering $v_1, \dots, v_{n-1}$ of the vertices in $U-u$, there exist
	disjoint sets $S_u^-, S_u^+ \subseteq U \setminus \{u\}$ that are a prefix and a suffix of the sequence such that
	\begin{align}
		|\partial(u, S_u^-)| = |\partial(u, S_u^+)| = \left\lceil \frac{|\partial(u)|}{2} \right\rceil - 1.
	\end{align}
	These sets can be found using $O(\log n)$ queries and $\tO(n)$ time.
\end{lemma}

\begin{proof}
	Let $L_j=\{v_1,v_2, \dots \dots, v_j\}$ and $R_j=\{v_j, v_{j+1}, \dots, v_{n-1}\}$ for $j\in [n-1]$. First, check if $\partial(v_1)>\partial(v_1+u)$ in which case, we can conclude $S_u^{-}=\emptyset, S_{u}^{+}=\{v_2,v_3, \dots, v_{n-1}\}$ because $\{u,v_1\}$ has to be the only edge using \Cref{obs:query_observation}. Similarly for $v_{n-1}$. So from now on, assume that $\partial(v_1)\leq \partial(v_1+u)$ and $\partial(v_{n-1})\leq \partial(v_{n-1}+u)$. Equivalently, this can also be written as $\partial(R_2)\geq \partial(R_2+u), \partial(L_{n-2})\geq \partial(L_{n-2}+u)$.
	
	In what follows, we will first describe how to compute $S_u^{-}$ and then use a similar argument to compute $S_u^+$. If $\partial(v_1)=\partial(v_1+u)$, then again we can simply conclude $S_{u}^{-}=\emptyset$ because this implies that $|\partial(u)|=2$ and $\{u, v_1\}$ is an edge. Otherwise, if $\partial(v_1)<\partial(v_1+u)$, we can search for $S_u^{-}$ as a tipping point for the chain defined by the sequence $v_1, v_2, \dots, v_{n-2}$ using \Cref{lem:cut_query_technical} because $\partial(v_1)<\partial(v_1+u)$ and $\partial(v_1, \dots, v_{n-2})\geq \partial(v_1, \dots, v_{n-2}+u)$. By the arguments
	in~\Cref{lem:cut_query_technical}, there exists $i\in [n-2]$ such that
	$v_i \in N(u)$ and
	\begin{align}
		\label{eqn:tipping-point}   |\partial(u, L_{i-1})| <
		\tfrac{1}{2}|\partial(u)|, \quad |\partial(u, L_i)| \geq
		\tfrac{1}{2}|\partial(u)|. 
	\end{align}
	This implies that $S_u^- := L_{i-1}$ satisfies
	$|\partial(u, S_u^-)| = \lceil \tfrac{|\partial(u)|}{2} \rceil - 1$.
	We can use the same argument to get $S_{u}^+$.
	Moreover, since $S_u^-$  and $S_u^+$ are prefix and suffix of the same
	ordering, and contain no more than half the neighbors of $u$, they
	must be disjoint. The proof makes two calls to
	\Cref{lem:cut_query_technical}, one for each ordering, with $O(n)$
	preprocessing. Each call takes $\tO(n)$ time, so the total running
	time is $\tO(n)$. 
\end{proof}

\subsection{Proof of Structural Primitives Lemma}

We can now prove the main structural result
\Cref{lem:cut-query-structural}, which we restate for convenience:

\CutStructureLemma*
Before we prove the theorem, we first prove a primitive that shows how to identify if a vetrex is isolated using constant queries when $n\geq 4$:
\begin{lemma}[Home Alone]
	\label{lem:vertex-home-alone}
	For any three vertices $u,v,w\in U$, vertex $u$ is isolated iff 
	\begin{align}
		|\partial(v)|=|\partial(u+v)|, |\partial(w)|=|\partial(u+w)|, |\partial(v+w)|=|\partial(u+v+w)|.
	\end{align}
\end{lemma}
\begin{proof}
	The three equalities are trivial when $u$ is isolated. Given the equalities, we observe that they can be written equivalently as 
	\begin{align}
		|\partial(u,v)|=\tfrac{1}{2}|\partial(u)|, |\partial(u,w)|=\tfrac{1}{2}|\partial(u)|, |\partial(u,v+w)|=\tfrac{1}{2}|\partial(u)|.
	\end{align}
	Using the fact that $|\partial(u,v+w)|=|\partial(u,v)|+ |\partial(u,w)|$ gives $|\partial(u)|=0$.
\end{proof}
\begin{proof}[Proof of \Cref{lem:cut-query-structural}]
	We first check if $\partial (u) =\emptyset$ using \Cref{lem:vertex-home-alone}. If $\partial (u) =\emptyset$, there are no neighbors of $u$ in $A$ in part~\ref{item:struct1} and we can certify that $N\cap T=\emptyset$ in part~\ref{item:struct2}. Assume $\partial (u)\neq \emptyset$ from here on.
	
	In proving part~\ref{item:struct1}, and determining which of
	$v_1, \dots, v_k \in U \setminus \{u\}$ are neighbors of $u$, we
	begin by computing the disjoint sets
	$S_u^-, S_u^+ \subseteq U \setminus \{u\}$ with exactly
	$\lceil \tfrac{|\partial(u)|}{2} \rceil - 1$ neighbors of $u$; this
	takes $O(\log n)$ queries and $\tO(n)$ time due to \Cref{lem:medians-u}.
	
	For each $v_i$ we choose $S$ to be one of $S_u^-$ or $S_u^+$ in
	which $v_i$ does not appear---say $S := S_u^+$. Since
	$|\partial(u, S)| < \tfrac{1}{2}|\partial(u)|$, we know from
	\Cref{obs:query_observation} that
	$\text{sign}(|\partial(S)| - |\partial(S + u)|) = -1.$ We then add
	$v_i$ to $S$ and ask for
	$\text{sign}(|\partial(S + v_i)| - |\partial(S + v_i + u)|)$.  Since
	$\partial(S, u)$ is just one edge short of reaching half the degree
	of $u$, this sign increases if and only if $v_i \in N(u)$.  Thus,
	each test requires one query, and the total query complexity is
	$k + O(\log n)$. After computing the median sets $S_u^-$ and $S_u^+$ in $\tO(n)$ time, we partition the vertices $v_1, \dots, v_k$ into two groups based on whether each $v_i$ lies in $S_u^-$ or $S_u^+$; this takes $O(k)$ time using bit vector lookups. For each group, we fix the corresponding set $S := S_u^+$ or $S_u^-$ and write its bit vector once in $O(n)$ time. Each query $(S + v_i, S + v_i + u)$ is then constructed in $O(1)$ time via bit flipping. The total running time is $\tO(n + k)$.
	
	This proves \Cref{lem:cut-query-structural}\ref{item:struct1}.
	
	\medskip
	For part~\ref{item:struct2} of the lemma, consider the ordered
	sequence $T = (v_1, \dots, v_t) \subseteq U \setminus \{u\}$ and
	extend it to a total ordering of $U \setminus \{u\}$ by appending
	the remaining vertices arbitrarily, forming the sequence
	$(v_1, \dots, v_t, v_{t+1}, \dots, v_{n-1})$.
	Define suffixes $R_k := \{v_k, \dots, v_{n-1}\}$ for each $k$; 
	by \Cref{lem:medians-u}, there exists some $k$ such that
	\begin{align}
		|\partial(u, R_k)| = \left\lceil \tfrac{|\partial(u)|}{2} \right\rceil - 1,
	\end{align}
	and such a suffix $S := R_k$ can be found using $O(\log n)$ queries and $\tO(n)$ time via \Cref{lem:medians-u}. To verify that $|\partial(u, T)| > 0$, 
	consider the sequence $(v_{t+1}, \dots, v_{n-1}, v_1, \dots, v_t)$ with $T$ placed all the way at the end. The suffix median of this ordering computed using \Cref{lem:medians-u} is the same as $S$ iff $|\partial(u, T)| > 0$. Otherwise, define the nested sequence
	\begin{align}
		S_0 &:= S, \\
		S_i &:= S \cup \{v_1, \dots, v_i\} \quad \text{for } i = 1, \dots, \min(t,k).
	\end{align}
	(Recall that $|T| = t$ and $|S| = n-k+1$.)  We have:
	\begin{align}
		|\partial(u, S_0)|&= |\partial(u, S)|=\left\lceil \tfrac{|\partial(u)|}{2} \right\rceil - 1, \\
		|\partial(u, S_{\min(t,k)})| &= |\partial(u, S)| + |\partial(u, T \setminus S)| \geq \tfrac{1}{2}|\partial(u)|\label{eqn:delta(u,T)}.
	\end{align}
	To justify the inequality in \eqref{eqn:delta(u,T)}: if
	$|\partial(u, T)| > 0$, then $|\partial(u, T \setminus S)| > 0$,
	because otherwise if $|\partial(u, T \setminus S)| = 0$, then
	$N(u) \subseteq S$. But this would imply
	$|\partial(u, S)| = |\partial(u)|$, contradicting the fact that less
	than half the edges incident to $u$ go into $S$ (by the choice of
	$S$). Hence, $T \setminus S$ must contain a neighbor of $u$.
	
	By \Cref{lem:cut_query_technical}, there exists an index $i$ such
	that the sign of the query $(S_{i-1}, S_{i-1} + u)$ is strictly
	smaller than that of $(S_i, S_i + u)$. The corresponding vertex
	$v_i$ is the first vertex in $T$ adjacent to $u$, and binary search
	over $i \in [\min(t,k)]$ finds it using $O(\log n)$ queries and $\tO(n)$ time.
\end{proof}

\EdgeExtraction*
\begin{proof}
	We assume that $|\partial(u, T)| \leq \left\lfloor \tfrac{1}{2}|\partial(u)| \right\rfloor + 1$; otherwise, we split $T$ into $T_+ := T \cap S_u^+$ and $T_- := T \cap S_u^-$ using the median sets from \Cref{lem:medians-u}, and recursively extract edges from each. Both subsets satisfy the same degree bound, and the total query and runtime complexity remains the same.
	
	Let $S$ be a median set disjoint from $T$, which exists by construction and can be found using $O(\log n)$ queries and $\tO(n)$ time. Let $T = \{v_1, \dots, v_t\}$, and define the nested sets $S_j := S \cup \{v_1, \dots, v_j\}$. We maintain two bit vectors: one for the fixed set $S$, and one for the growing prefix of $T$.
	
	In each iteration, we find the smallest index $j$ such that the sign of the query $(S_{j-1}, S_{j-1} + u)$ is strictly smaller than that of $(S_j, S_j + u)$, indicating that $v_j \in N(u)$. We locate this index using the doubling version of binary search. Once $v_j$ is found, we report the edge $\{u, v_j\}$, remove $v_1, \dots, v_j$ from $T$, and repeat. Since we extract at most $r$ edges, we perform at most $r$ such searches, each requiring $O(\log n)$ queries, for a total of $O(r \log n)$ queries overall.
	
	\paragraph{Running time.}
	Constructing the median set $S$ takes $\tO(n)$ time. To construct the query sets, we reuse the bit vector for $S$ and build each query by appending a prefix $\{v_1, \dots, v_i\} \subseteq T$. In each iteration, the doubling search explores prefixes of geometrically increasing size until it finds a neighbor at index $j$. The total size of all prefixes queried in that iteration is $O(j \log t)$, so the time spent constructing queries is $O(j \log t)$. Deleting $v_1, \dots, v_j$ from $T$ also takes $\tO(j)$ time. The values $j_1, \dots, j_r$ across all $r$ iterations sum to at most $n$ as these are gaps between the edge indices, so the overall total time taken is $O\left(\sum_{i=1}^r j_i \log n\right) = \tO(n)$.
\end{proof}

\subsection{Cut sparsifiers using $\widetilde{O}(n)$ queries}
\label{sec:cut-sparsifier-details}


Using \Cref{lem:contracted-graphs}, we show that we can implement the
sparsifier construction from \cite[Algorithm
3.4]{RSW18-value-oracle-mincuts} and the min-cut computation from
\cite[Algorithm 4.1]{RSW18-value-oracle-mincuts} using
$\widetilde{O}(n/\eps^2)$ cut comparison queries. The mentioned
algorithms and relevant theorems from
\cite{RSW18-value-oracle-mincuts} are mentioned verbatim below:


\begin{algorithm}[H]
	\caption{(Approximating Edge Strengths and Sampling a Sparsifier $H$)}
	\label{alg:approx-edge-strengths}
	
	\textbf{Input:} An accuracy parameter $\varepsilon$, and a cut-query oracle for graph $G$.
	
	\begin{enumerate}
		\item Initialize an empty graph $H$ on $n$ vertices and $G_0\leftarrow G$.
		
		\item For $j = 0, \ldots, \log n$, set $\kappa_j = n2^{-j}$ and:
		\begin{enumerate}
			\item Subsample $G'_j$ from $G_j$ by taking each edge of $G_j$ with probability 
			\[
			q_j = \min\left(\frac{100 \cdot 40 \cdot \ln n}{\kappa_j}, 1\right).
			\]
			
			\item In each connected component of $G'_j$:
			\begin{enumerate}
				\item While there exists a cut of size $\leq q_j \cdot \frac{4}{5}\kappa_j$, remove the edges from the cut. Let the connected components induced by removing the cut edges be $C_1, \ldots, C_r$.
				
				\item For every $i \in [r]$ and every edge (known or unknown) with both endpoints in $C_i$, set the approximate edge strength $k'_e := \frac{1}{2\kappa_j}$  
				(alternatively, subsample every edge in $C_i \times C_i$ with probability $2q_j/\varepsilon^2$ and add it to $H$ with weight $\varepsilon^2 / 2q_j$).
				
				\item Set $G_{j+1}$ as the graph obtained by contracting $C_i$ for each $i \in [r]$ in $G_j$.
			\end{enumerate}
		\end{enumerate}
	\end{enumerate}
	
	\textbf{Output:} The edge strength approximators $\{k'_e\}$ (or the sparsifier $H$).
\end{algorithm}

\begin{theorem}[Theorem 3.5 from \cite{RSW18-value-oracle-mincuts}]
	\label{thm:edge-strength-sparsifier}
	For each edge $e \in G$, the approximate edge strength given in Algorithm~\ref{alg:approx-edge-strengths} satisfies $\frac{1}{4}k_e \leq k'_e \leq k_e$. Furthermore, the algorithm requires $\widetilde{O}(n / \varepsilon^2)$ oracle queries to produce a sparsifier $H$ with the following properties:
	\begin{itemize}[nosep]
		\item $H$ has $O(n \log n / \varepsilon^2)$ edges.
		\item The maximum weight of any edge $e$ in $H$ is $O(\varepsilon^2 k_e / \log n)$.
		\item Every cut in $H$ approximates the corresponding cut in $G$ within a $(1 \pm \varepsilon)$ factor.
	\end{itemize}
\end{theorem}

\begin{algorithm}
	\captionsetup{labelformat=empty}
	\caption{Simpler global Min Cut with $\tO(n)$ oracle queries}
	\label{alg:simpler-mincut}
	
	\textbf{Input:} Oracle access to the cut values of an unweighted simple graph $G$.
	
	\begin{enumerate}[label=\arabic*.]
		\item Compute all of the single-vertex cuts.
		
		\item Compute a sparsifier $H$ of $G$ using Algorithm~\ref{alg:approx-edge-strengths} with small constant $\varepsilon$.
		
		\item Find all non-singleton cuts of size at most $(1 + 3\varepsilon)$ times the size of the minimum cut in $H$, and contract any edge which is not in such a cut. Call the resulting graph $G'$.
		
		\item If the number of edges between the super-vertices of $G'$ is $O(n)$, learn all of the edges between the super-vertices of $G'$ and compute the minimum cut.
	\end{enumerate}
	
	\textbf{Output:} Return the best cut seen over the course of the algorithm.
\end{algorithm}

\begin{theorem}[Theorem 4.2 from \cite{RSW18-value-oracle-mincuts}]
	\label{thm:mincut-simpler}
	Algorithm~\ref{alg:simpler-mincut} uses $\widetilde{O}(n)$ queries and finds the exact minimum cut in $G$ with high probability.
\end{theorem}

\begin{proof}[Proof of \Cref{thm:unweighted-cuts}]
	For each $0\leq j\leq \log n$, step 2(a) of Algorithm~\ref{alg:approx-edge-strengths} can be implemented using \Cref{lem:contracted-graphs} with  $O(|E_j|\cdot q_j\cdot \log n)+\tO(n)$ comparison queries in expectation; which is shown to be $\widetilde{O}(n)$ in the proof of \Cref{thm:edge-strength-sparsifier}. The runtime bound of the same step is $\tO(n^2)$. Step 2(b)(ii) of Algorithm~\ref{alg:approx-edge-strengths} also takes $\widetilde{O}(n/\eps^2)$ queries and $\tO(n^2)$ time because the number of edges sampled and added to $H$ is shown to be $\widetilde{O}(n/\eps^2)$ (see proof of Theorem 4.2, \cite{RSW18-value-oracle-mincuts}) and this sub-sampling can be done again using \Cref{lem:contracted-graphs}. In Algorithm~\ref{alg:simpler-mincut}, since step (1) is only used to compare the singleton cuts to output the minimum cut seen by the algorithm, we can implement this using comparison in ${O}(n)$ queries and time. In step (3), all the approximate cuts can be computed in $\tO(n^2)$ time using \cite{HHS24-mincut-cactus}. In step (4), all the edges of $G'$ can be learned using \Cref{lem:contracted-graphs} with $p=1$. The proof of \cref{thm:mincut-simpler} shows that the number of edges in $G'$ is at most $O(n)$ with high probability. This implies that all the steps in the algorithms take $\widetilde{O}(n/\eps^2)$ queries and $\tO(n^2)$ time. There are at most $\log n$ steps in any algorithm which proves the desired bounds. 
\end{proof}
\subsection{Weighted Min-Cut: Special Cases and Stronger Oracles}
\label{sec:weighted-cuts}

In the weighted setting, where the edges of the graph $G = (V, E)$ have unknown non-negative weights, the techniques developed for the unweighted case (where weights are $0$ or $1$; see \Cref{sec:minimum-cut}) no longer apply. Even when the weights are restricted to integers in $[0, B]$, little is known.

\subsubsection{Weighted regular graphs}
We consider the case when the graph satisfies the \emph{weighted regularity} condition, i.e., $|w^*(\partial u)|$ is identical for all $u \in V$. Under this assumption,
\begin{align}
\label{eqn:edge-weight-extraction}
   w^*(\{u, v\}) = \tfrac{1}{2}\big(w^*(\partial u) + w^*(\partial v) - w^*(\partial \{u, v\})\big), 
\end{align} 
and hence the relative order of edge weights $w^*_{\{u,v\}}$ is simply the reverse of the order of the cuts $\partial\{u,v\}$ sorted by their cut values. Since these comparisons can be performed using only cut queries, we can recover the rank order of all edge weights.

Because the weights are bounded by $B$, there are at most $\ell := n^2B + 1$ distinct cut values, corresponding to $\ell$ ordered \emph{buckets} of edges. We wish to classify these buckets into $O(\log B)$ weight ranges:
\[
0, [2^{0}, 2^{1}), [2^{1}, 2^{2}), \ldots, [2^{\lfloor \log_2 B \rfloor}, 2^{\lfloor \log_2 B \rfloor + 1}).
\]
A naive assignment of each bucket to one of these $O(\log B)$ classes would yield $(O(\log B))^{\ell}$ possibilities. However, since the buckets are ordered, we can choose the $O(\log B)$ boundary points separating consecutive classes along this sequence. Thus, the number of possible partitions is $O(\ell^{O(\log B)})$,
and one such partition corresponds to a $2$-approximation of the true edge weights.

Using these approximate weights, we enumerate all $2$-approximate cuts. By Karger’s result, there are only $n^{O(1)}$ such cuts, and these can be enumerated in $\tO(n^2)$ time (see \cite{HHS24-mincut-cactus}). Since our guessed weights form a $2$-approximation, the true minimum cut (which is a $2$-approximate cut under the guessed weights) appears among them. Comparing all enumerated cuts across all weight guesses and returning the smallest yields a total running time and number of cut comparisons of $O((n B)^{O(\log B)})$.

The guessing step can be extended to the more general case where there are $r$ distinct weighted degrees. In this case, we first classify the vertices into $V_1, \dots, V_r$ according to their degrees (this can be done by comparing their degree cuts). Each edge then lies between some pair $(V_i, V_j)$, and we accordingly classify all edges into $r + \binom{r}{2}$ classes $E_{\{i,j\}}$. 

Observe that for two edges $\{u_1, v_1\}, \{u_2, v_2\} \in E_{\{i,j\}}$ within the same class, their weights can be compared by comparing the cuts $\partial\{u_1, v_1\}$ and $\partial\{u_2, v_2\}$ using Equation \eqref{eqn:edge-weight-extraction}. Hence, we apply the same guessing procedure as before independently within each class, resulting in a total of $O(\ell^{O(r^2 \log B)})$ guesses. We can then proceed exactly as in the uniform-degree case. Consequently, the overall running time and number of cut comparisons are $O((nB)^{O(r^2 \log B)})$.
\subsubsection{Stronger Oracles}

An alternative to imposing structural assumptions is to consider stronger oracle models. We highlight two natural extensions:

\begin{enumerate}
    \item \textbf{Comparing marginals:} Instead of comparing the objective values of feasible sets, i.e., $w(S)$ and $w(T)$ for $S, T \in \cF$, we allow comparisons between \emph{marginal differences} such as $w(S) - w(S')$ and $w(T) - w(T')$ for $S, S', T, T' \in \cF$.
    \item \textbf{Comparing arbitrary sets:} We relax the restriction that comparisons must involve only feasible sets, and instead allow comparing $w(S)$ and $w(T)$ for any $S, T \subseteq U$, while the optimization goal remains finding $\arg\min_{S \in \cF} w(S)$.
\end{enumerate}

In both of these stronger models, the minimum cut can be found efficiently, even in the weighted setting with arbitrary non-negative weights. The key observation is that the Nagamochi--Ibaraki algorithm~\cite{NagamochiI}  for minimum cuts only requires comparisons of \emph{differences} of cut values. More specifically, at each step it only needs to find
\[
    \arg\min_{u \in V \setminus S} \big(w(\partial(S + u)) - w(\partial u)\big),
\]
where the graph at this point is a contraction of the original, so the current vertex set $V$ may consist of disjoint subsets of the original vertices (which poses no issue). The algorithm then outputs the best cut among a small collection of candidate cuts, which can be selected using cut comparisons.

Notably, the identity
\[
    \arg\min_{u \in V \setminus S} \big(w(\partial(S + u)) - w(\partial u)\big)
    = \arg\min_{u \in V \setminus S} w(\partial(S, u))
\]
implies that the Nagamochi--Ibaraki algorithm can also be implemented under the second, arbitrary-set comparison model. Finally, the Nagamochi--Ibaraki algorithm has a natural extension to general non-negative symmetric submodular functions, as shown by \cite{Queyranne98}.

\subsection{Illustrative Examples}
\label{sec:useful-examples}

\subsubsection{Separations between Comparison and Value Queries}
\label{sec:non-adaptive}

In the weighted setting, consider taking two disjoint cycles $C_1$ and
$C_2$ of $k = n/2$ vertices and with unit-weight edges, and adding in
a perfect matching of size $k$ between them, with edges of cost
$\eps$. This is the \emph{circular ladder graph}, with cross-edge
weight $\eps$.

\begin{claim}
	\label{clm:small-queries}
	There exist non-adaptive algorithms making $O(n^2)$ value queries to
	sets of size one and two that can reconstruct any graph, but for any
	non-adaptive algorithm in the comparison-query model making queries
	to sets of constant size (or even $\ll n/2$ size), there exist
	weighted $n$-vertex graphs $G^+, G^-$ which cannot find the minimum
	cut.
\end{claim}

\begin{proof}
	Take the circular ladder graph with
	$\eps = \frac{2}{k-1} \pm \gamma$ for some tiny $\gamma > 0$; the
	sign in front of the $\gamma$ will determine whether the min-cut is
	a single-vertex cut with cost $2+\eps$, or the bisection
	$(C_1, C_2)$ with cost $k\eps$. However, we cannot distinguish these
	two cases if we compare sets $S, T$ of size $\ll n/2$. (This is true
	even if we know the edges of the graph and the entire construction,
	except the sign in front of $\gamma$. This is in contrast to the
	value query setting, where asking the value of $f(S)$ for sets of
	size $1$ and $2$ allows us to completely reconstruct the graph.
\end{proof}

\begin{claim}
	\label{clm:expo-non-adapt}
	There exist non-adaptive algorithms making $O(n^2)$ value queries to
	reconstruct the graph, but for any non-adaptive algorithm in the
	comparison-query model, there exist weighted $n$-vertex graphs
	$G^+, G^-$ which require $\exp(n)$ comparison queries to have high
	success probability.  
\end{claim}

\begin{proof}
	Again, take the circular ladder graph, and rename the vertices
	uniformly at random. Assume that the cross-edge weight is
	$\eps = \frac{2}{k-1} \pm \gamma$ for some tiny $\gamma > 0$; the
	sign in front of the $\gamma$ will determine whether the degree cut
	or the partition $(C_1, C_2)$ is optimal. Again, for value queries
	we can non-adaptively reconstruct the graph using $O(n^2)$
	queries. But for comparison queries, the only way to learn the sign
	in front of $\gamma$ would be to query the correct partition with a
	degree cut. Now if the queries are non-adaptive, and the vertices of
	the graph have been uniformly and randomly renumbered, the number of
	non-adaptive queries would have to be exponential in $n$ in order to
	have a high probability of success. 
\end{proof}

\subsubsection{Non-Reconstructable Graphs}
\label{sec:non-reconstr-graphs}

The unweighted graph pairs $(K_2, \bar{K_2})$ and $(K_3, \bar{K_3})$
are not distinguishable, since essentially the only feasible sets $S$
are singleton sets (recall that by symmetry, $f(S) = f(\bar{S})$), and
all of them have the same cut value. However, as we showed in
\Cref{thm:graph_recovery}, we can recover all unweighted simple graphs other than $K_2, \bar{K_2},K_3, \bar{K_3} $ using comparison queries. 

In contrast, many connected weighted graphs cannot be recovered using
just comparison queries. This is not surprising: we cannot distinguish
between a graph with weights $w_e$, and those with weight $\alpha w_e$
for all $\alpha > 0$. But there are other kinds of barriers apart from
a simple scaling of weights---here is an example. Given any graph
$G = (V,E)$ with unit weight edges, add another spanning tree
$T = (V,F)$ on the same vertices, and give the $i^{th}$ edge
$e_i \in F$ a weight of $w_T(e_i) = n^2 \cdot 2^i$. This weight
function ensures that the $F$-weights of all cuts are
distinct. Moreover, for a set $S \sse V$,
$w(\partial S) = |\partial_G S| + w_T(\partial_T S)$. Since the $w_T$
weights are scaled by $n^2$ (and hence much larger than the
contribution due to the unit-weight edges), we know that
\[ w(\partial S) > w(\partial S') \iff w_T(\partial_T S) >
w_T(\partial_T S) \]
and hence we cannot infer the structure of the unit-weight edges.

\subsubsection{Finding Heaviest Edge Incident to a Vertex}\label{sec:heavy}

We can show that comparison queries do not allow us to identify, for a vertex $v$, the heaviest edge adjacent to it. Finding this edge would be useful, e.g., in the min-cut algorithm of \cite{SW94}. Indeed, consider the 4-vertex graph $\{a, b, c, d\}$ and the 6 edge weight variables. Here are two universes:

\begin{center}
	\begin{tabular}{|>{\centering\arraybackslash}m{0.2\textwidth}|>{\centering\arraybackslash}m{0.3\textwidth}|>{\centering\arraybackslash}m{0.3\textwidth}|}
		\hline
		\textbf{Edge Weight} & \textbf{Universe 1 (\(w_{ab} > w_{ac}\))} & \textbf{Universe 2 (\(w_{ab} < w_{ac}\))} \\
		\hline
		\textbf{\(w_{ab}\)} & \textbf{1000.01} & \textbf{999.99} \\
		\textbf{\(w_{ac}\)} & \textbf{999.99} & \textbf{1000.01} \\
		\textbf{\(w_{ad}\)} & \textbf{10} & \textbf{10} \\
		\(w_{bc}\) & 100 & 100 \\
		\(w_{bd}\) & 50 & 50 \\
		\(w_{cd}\) & 1 & 1 \\
		\hline
	\end{tabular}
\end{center}

In both scenarios, \(w_{ad} = 10\) is much less than \(w_{ab}\) and \(w_{ac}\) (which are both \(\sim\)1000). However, a calculation shows that the order of the cuts is \textbf{identical} in both universes:
\[ f(\{a,d\}) > f(\{a\}) > f(\{a,b\}) > f(\{b\}) > f(\{a,c\}) > f(\{c\}) > f(\{d\}). \]
Hence, given this specific ordering of cuts, we cannot distinguish between Case~1 (where $(a,b)$ is the max-weight edge adjacent to $a$) or Case~2 (where $(a,c)$ is). 

\section{Matroids, Matchings, and Paths}
\label{sec:comb-algorithms}

In this section, we give efficient comparison-based algorithms to optimize over matroids, matroid intersections, and shortest paths.

\subsection{Matroid Bases}
\label{sec:matroid-bases}

We now give our algorithm to compute a minimum-weight basis of a
matroid $\cM=(\elts,\cI)$ with $|\elts| = n$ elements, where we are only
allowed to compare matroid bases.

Our
main result for matroid bases is the following:

\begin{restatable}[Matroid Bases]{theorem}{Matroid}
  \label{thm:matroid-bases}
  There is an algorithm outputs the minimum-weight basis of a matroid
  on $n$ elements using $O(n \log n)$ comparison queries in $\tO(n^2)$
  time.
\end{restatable}


\begin{proof}
	First, let us consider graphical matroids, where the bases are
	spanning trees of a graph. Given any graph with edge set $E$, we can
	first partition the graph into its $2$-vertex-connected components
	$E_1, \ldots, E_k$ (note that these are the edge sets of the
	components); the edge set of the minimum-weight spanning tree (MST)
	for $E$ is the union of those for the $E_i$'s, so it suffices to
	focus on a $2$-connected component induced by the edges in some
	$E_i$. We claim we can simulate Kruskal's algorithm for MST in each of these components, which
	just compares weights of edges: indeed, if this algorithm compares
	edges $e, e' \in E_i$, we can find a base $B$ containing $e$, such
	that $B' := (B \setminus e) \cup e'$ is also a base. (We defer the
	proof, since it follows from the result below.) Now comparing $e,e'$
	has the same answer as comparing $B,B'$.
	
	The same argument holds for arbitrary matroids, where we recall that the notion of $2$-vertex connectivity in graphs extends to matroids~\cite{oxley-matroids}:
	\begin{lemma}
		\label{lem:matroid}
		For any matroid $\cM=(\cI,\elts)$, let $\cB(\cM)$ and $\cC(\cM)$
		be the collection of all bases and circuits in the matroid.
		\begin{enumerate}[nosep]
			\item If elements $e,e'\in E$ share a circuit $C\in \cC(\cM)$,
			then there exist bases $B, B' \in \cB(\cM)$ such that
			$B'=B+e'- e$. Call such pair of elements $e,e'$
			\textbf{comparable}.
			\item \label{item:eq-class} The binary relation $\gamma_{\cM}$ on the elements of the
			matroid such that $(e,e')\in \gamma_{\cM}$ iff either $e=e'$ or
			$e,e'$ are comparable is an equivalance relation.
		\end{enumerate}
	\end{lemma}
    \begin{proof}
	For the first part, let $B$ be the basis that extends the
	independent set $C-e'$. This means that the fundamental circuit in
	$B+e'$ is $C$ implying that $B':=B+e'-e$ is also a basis. The second
	part is proved in \cite[Proposition 4.1.2]{oxley-matroids}. The
	equivalence classes in \cref{lem:matroid}(\ref{item:eq-class}) are
	called the \emph{connected components} of $\cM$.
\end{proof}
	Now we can run the analog of Kruskal's greedy algorithm for matroids
	on the connected components of $\cM$; this uses
	$O(|\elts_i| \log |\elts_i|)$ element (and hence base) comparisons for each
	component; summing over all components gives us $O(|\elts| \log |\elts|)$,
	as claimed. 
	
	We now give details on efficient algorithms for decomposition of the matroid into connected components and obtaining two bases $B, B'$ such that $B\Delta B'=\{x,y\}$ for every pair  $x,y$ compared by the algorithm, and overall running time bounds:
		The running time consists of the time taken to decompose the matroid into its connected components, and obtaining two bases $B, B'$ such that $B\Delta B'=\{x,y\}$ for every pair  $x,y$ compared by the algorithm. For this, we use the results from \cite{krogdahl-basis-graph} that connect the \emph{basis exchange graph} with the connected components of the matroid. The following are a definition and a lemma:
	
	For any basis $B$ of $\cM$, the basis exchange graph $H(B)$ is defined as the bipartite graph with color classes $B$ and $U\backslash B$. For $x\notin B, y\in B$, we have $\{x,y\}\in H$ iff $B+x-y$ is also a basis. The following lemma connects this exchange graph with matroid connectivity:
	\begin{lemma}[Section 6, \cite{krogdahl-basis-graph}]
		\label{lem:basis-connect}
		For any basis $B\in \cM$, 
		\begin{enumerate}[nosep]
			\item \label{item:basis-connect1} The nodes in the connected components (in the graphic sense) of $H(B)$ are the connected components (in the matroid sense) of matroid $\cM$. 
			\item \label{item:basis-connect2} For $x,y$ in the same connected component of $H(B)$ (or equivalently $\cM$) such that $x\notin B$ and $y\in B$, if $P$ is the shortest path connecting $x$ and $y$ in $H(B)$, then $B':=B\Delta P$ is a basis such that $B'+y-x$ is a basis. 
		\end{enumerate}
	\end{lemma}
	Given \Cref{lem:basis-connect}, the matroid decompostion step requires $O(n^2)$ matroid independence calls (in fact, the calls are always about whether a set is a basis or not) to build the exchange graph using \Cref{lem:basis-connect} part \ref{item:basis-connect1}  and $O(n^2)$ time to identify connected components.  
	
	Next, to simulate any pairwise comparison between elements $x,y$, we obtain the two bases $B\ni x, B'\ni y$ such that $B\Delta B'=\{x,y\}$ using part \ref{item:basis-connect2} of \Cref{lem:basis-connect}. The implementation of Kruskal's algorithm requires sorting the elements in each connected component which requires $O(c_i\log c_i)$ comparisons in each component with $c_i$ elements. Each such comparison takes $O(n)$ time to find the path and write down the bases to input to the comparison oracle. Adding everything, we get the $O(n^2\log n)$ time bound as claimed. 
\end{proof}

\subsection{Matroid Intersection}
\label{sec:matroid-intersection}

We now turn to finding \emph{common independent sets}
$ S \in \mathcal{I}_1 \cap \mathcal{I}_2 $ between two matroids
$ \mathcal{M}_1 = (U, \mathcal{I}_1) $ and
$ \mathcal{M}_2 = (U, \mathcal{I}_2) $.  Our main result in this case
is the following:

\begin{restatable}[Matroid Intersection]{theorem}{MatroidInt}
	\label{thm:matroid-intersection}
	Let $\mathcal{M}_1 = (U, \mathcal{I}_1)$ and
	$\mathcal{M}_2 = (U, \mathcal{I}_2)$ be two matroids 
	defined on a ground set $U$ containing $n$ elements.
	Then, there is an algorithm that outputs the minimum-weight set that is in both $\mathcal{I}_1, \mathcal{I}_2$ using $O(n^4)$
	comparison queries in $O(n^4)$ time.
\end{restatable}


To prove this, we show that Edmond's classical weighted matroid
intersection algorithm, as presented
in~\cite[Section~41.3]{Schrijver-book-combi-opt}, can be implemented
using only comparisons of common independent sets. As in the previous
section, we first consider the simpler bipartite matching case, where
the shortest augmenting paths algorithm constructs extremal matchings
via shortest paths in an exchange graph. Our key observation is that
these shortest-path steps---e.g., via Bellman–Ford---can themselves be
implemented using only comparisons between matchings.

Formally, let $G = (V, E)$ be a bipartite graph with color classes
$V = L \cup R$. Given an \emph{extremal matching} $M$ of size~$k$
(i.e., a min-cost matching with $k$ edges), we describe a primitive to
obtain an extremal matching $M'$ of size~$k+1$.

The \emph{$M$-exchange graph} $D$ orients matching edges from $R$ to
$L$ and non-matching edges from $L$ to $R$ in $G$. Let $L'$ and $R'$
be the unmatched vertices in $L$ and $R$. An \emph{$M$-augmenting
	path} is a directed path in $D$ from a vertex in $L'$ to one in
$R'$. Define the edge length function $\ell : E \to \mathbb{R}$ as:
$\ell(e) := -w(e)$ if $e \in M$, and $w(e)$ otherwise. The following
result is standard (see, e.g., Theorem 3.5, Proposition 1 in Section 3.5 of \cite{combiopt_schrijver}). 

\begin{lemma}
	\label{lem:augmenting-paths}
	Let $M$ be an extremal matching. Then:
	\begin{enumerate}[nosep]
		\item The exchange graph $D$ has no negative-length cycles.
		\item If $P$ is a minimum-length $M$-augmenting path, then $M' := M \triangle P$ is also extremal.
	\end{enumerate}
\end{lemma}

By iteratively applying \Cref{lem:augmenting-paths}, we construct
extremal matchings of increasing size, from size~$0$ to $n/2$, and
return the one with minimum total weight. Moreover, to implement this
using only comparisons between matchings, we describe a variant of the
Bellman-Ford algorithm. For each $u \in L$, let $p_u^{(k)}$ be the
shortest $M$-alternating path from $L'$ to $u$ using at most $k$
matching edges. For matched $u \in L$, let $u' \in R$ be its partner.
We initialize:
\[
p_u^{(1)} = 
\begin{cases}
	\emptyset & \text{if } u \in L', \\
	s \rightarrow u' \rightarrow u & \text{for } s = \arg\min_{s \in L'} \ell(s \rightarrow u' \rightarrow u).
\end{cases}
\]
Moreover, we recursively define
$p_u^{(k+1)} = \min_{v \in L} \big( p_u^{(k)},\; p_v^{(k)}
\rightarrow u' \rightarrow u \big)$, where the $\min$ operator
returns the path with minimum total length $\ell(p)$, ties broken
arbitrarily. Each such path $P$ yields a matching $M \triangle P$, and
satisfies $\ell(P) = w(M \triangle P)-w(M)$, so comparing path
lengths reduces to comparing weights of the resulting matchings.

To reach a free vertex $t \in R'$, compute:
$p_t = \min_{v \in L} \big( p_v^{(n)} \rightarrow t \big)$, which
again defines an alternating path. Hence, the full algorithm can be
implemented using only comparisons between matchings.

This insight extends naturally to the matroid intersection setting,
where shortest augmenting paths must also be minimal (i.e., having
fewest arcs among paths of equal weight), a property standard
algorithms like Bellman–Ford already ensure. We show that the
path-finding steps can again be carried out using only comparisons of
common independent sets. The details appear below:

\subsection{Generalization to Matroid Intersection}
\label{sec:matroid-intersect-appendix}

We now extend the concepts from bipartite matching to matroid
intersection. Let $\cM_1 = (E, \cI_1)$ and $\cM_2 = (E, \cI_2)$ be
matroids over a common ground set $E$, and let
$Y \in \cI_1 \cap \cI_2$ be a common independent set.
Define the \emph{exchange graph} $H(\cM_1, \cM_2, Y)$ as a directed
bipartite graph with color classes $Y$ and $E \setminus Y$. For
$y \in Y$ and $x \in E \setminus Y$:
\[
\begin{aligned}
	& (y,x) \in H \iff Y - y + x \in \cI_1, \\
	& (x,y) \in H \iff Y - y + x \in \cI_2.
\end{aligned}
\]
Let $H(\cM_1, Y), H(\cM_2,Y)$ be the subgraphs of $H$ with edges corresponding to $\cM_1, \cM_2$ respectively. 

We know the following lemma (see  \cite[Section 10.4]{combiopt_schrijver}):
\begin{lemma}[Matching lemmas]
	\label{lem:matching-matroid}
	For $Y\in \cI_1$, let $H_1=H(\cM_1, Y)$. 
	\begin{enumerate}[nosep,label=(\alph*)]
		\item \label{item:matching-matroid1} If $Z\in \cI_1$ with $|Y|=|Z|$, then $H_1$ contains a perfect matching on $Y\Delta Z$.
		\item \label{item:matching-matroid2} If $Z\subseteq E$ such that $H_1$ contains a unique perfect matching on $Y\Delta Z$, then $Z\in \cI_1$.
	\end{enumerate}
\end{lemma}

Let $X_1 := \{x \in E \setminus Y : Y + x \in \cI_1\}$ and $X_2 := \{x \in E \setminus Y : Y + x \in \cI_2\}$.
Given a weight function $w : E \to \RR$, define the length of any path (or cycle) $P$ in $H$ by:
\[
\ell(P) := -\sum_{e \in P \cap Y} w_e + \sum_{e \in P \setminus Y} w_e.
\]
A common independent set $Y\in \cI_1\cap \cI_2$ is \emph{extreme} if
it minimizes weight among all common independent sets of the same
size. We know the following lemmas (see Theorem 10.7, Theorem 10.11, Theorem 10.12 in \cite{combiopt_schrijver})
\begin{lemma}
	\label{lem:termination}
	The subset $Y\in \cI_1\cap \cI_2$, is a maximum cardinality common independent set iff there is no directed path from $X_1$ to $X_2$ in $H$.
\end{lemma}

\begin{lemma}
	\label{lem:extreme-intersection}
	Consider a common independent set  $Y\in \cI_1\cap \cI_2$.
	\begin{enumerate}[nosep,label=(\roman*)]
		\item \label{item:extreme-intersection1} $Y$ is extreme if and only if $H(\cM_1, \cM_2, Y)$ contains no directed cycle of negative length.
		\item \label{item:extreme-intersection2} If $Y$ is extreme, and $P$ is a shortest (minimal-hop) path from $X_1$ to $X_2$ in $H$, then $Y \triangle P$ is also extreme.
	\end{enumerate}  
\end{lemma}
To mimic the Bellman--Ford algorithm, we generalize the notion of augmenting paths using \emph{minimal paths and cycles}, defined with respect to both weight and hop-count.

\paragraph{Minimal Cycles and Paths.}
A cycle $C$ in $H$ is \emph{minimal} if no proper subcycle $C'$ satisfies $\ell(C') \leq \ell(C)$. Similarly, for $y \in Y$, define $\cP(y,t)$ to be the set of directed paths from $X_1$ to $y$ in $H$ using at most $2t-1$ arcs. A path $P \in \cP(y,t)$ is \emph{minimal} if it is lexicographically minimal with respect to length and number of arcs:
\[
\text{For all } P' \in \cP(y,t),\quad (\ell(P'), |P'|) \succeq (\ell(P), |P|).
\]
In the following lemma, we prove our main observation that augmenting an extreme common independent by a minimal path gives a common independent set.  
\begin{lemma}[Minimal paths and cycles]
	\label{lem:minimal-path-cycle}
	If $Y$ is extreme:
	\begin{enumerate}[nosep]
		\item For any minimal cycle $C$ in $H$, we have $Y \triangle C \in \cI_1 \cap \cI_2$.
		\item For any minimal path $P \in \cP(y,t)$, we have $Y \triangle P \in \cI_1 \cap \cI_2$.
	\end{enumerate}
\end{lemma}

\begin{proof}[Proof of \Cref{lem:minimal-path-cycle}]
	For the proof of part~(1), suppose $Z:=Y\Delta C \notin \cI_1\cap \cI_2$, assume by symmetry that $Z\notin \cI_1$. Let $N,M$ be the matching edges in $C$ corresponding to $\cM_1$ and $\cM_2$ respectively. Using \Cref{lem:matching-matroid}\ref{item:matching-matroid1}, we know that there exists a different matching $N'\neq N$ between the vertices $VC\cap Y$ and $VC\backslash Y$ with edges corresponding to $\cM_1$.
	
	Consider the multiset union of edges $N\cup N' \cup 2M$ which takes two copies of every edge from $M$ and a copy of every edge from $N$ and $N'$. Since these set of edges enter and leave every vertex exactly twice, the set of edges can be decomposed into a collection of cycles $C_1,\dots,C_p$ for some $p\geq 2$ such that some of the cycles have strictly fewer vertices than $C$. We also have $\sum_{i\in p}\ell(C_i)=2\ell(C)$. Using \Cref{lem:extreme-intersection}\ref{item:extreme-intersection1}, we know $\ell(C_i)\geq 0$ and since $p\geq 2$, there exists a cycle $C_i$ with lesser weight or the same weight with lesser number of vertices as $C$ contradicting the minimality of $C$. 
	
	The proof of part~(2) proceeds in a few steps:
	\begin{enumerate}[label=(2\alph*)]
		\item (\textbf{No re-entry}) \label{proof:lem-structural-2a} First, we prove that a minimal path in $\cP(y,t)$ never re-enters $X_1$ again. 
		If $P=(x_0, y_1,x_1, \dots, y_k, x_k, \dots, x_{t-1}, y_t)$ with $x_k\in X_1$, then observe that there is an edge from $y_k$ to $x_0$ because $Y+x_0\in \cI_1$ as $x_0\in \cI_1$ which implies $Y+x_0-y_k\in \cI_1$ and hence the edge. If we let $C=(x_0, y_1,x_1, \dots, y_k, x_0)$, and $P'=(x_k, y_{k+1}, \dots, x_{t-1}, y_t)$, we have $\ell(P)=\ell(C)+\ell(P')$ implying that $P'$ has smaller length and lesser number of arcs than $P$ as $\ell(C)\geq 0$ from \Cref{lem:extreme-intersection}\ref{item:extreme-intersection1} contradicting the minimality of $P$. An identical argument can be used to show that vertices are not revisited again in minimal paths.
		
		\item ($Y\Delta P \in \cI_1$) \label{proof:lem-structural-2b}  Next, we show that $Y\Delta P \in \cI_1$. Let $Z=\{y_1,x_1,\dots, y_{t-1},x_{t-1}\}$. We first show that $Y\Delta Z\in \cI_1$. Let $N$ be the set of matching edges $(y_i, x_i)$ corresponding to $\cM_1$ for $1\leq i\leq t-1$ and $M$ be the set of matching edges $(x_{j-1}, y_j)$ corresponding to $\cM_2$ for $1\leq j\leq t$. 
		If $Y\Delta Z\notin \cI_1$, then there is a different matching $N'$ between $Z\backslash Y$ and $Z\cap Y$ using \Cref{lem:matching-matroid}\ref{item:matching-matroid1}. Consider the multiset union of edges $N\cup N'\cup 2M\cup 2(y_k, x_0)$. 
		
		Since these set of edges enter and leave every vertex exactly twice, the set of edges can be decomposed into a collection of cycles $C_1,\dots,C_p$ for some $p\geq 2$ such that some of the cycles have strictly fewer vertices than $C$. We also have $\sumL_{i\in p}\ell(C_i)=2\ell(C)$. Using \Cref{lem:extreme-intersection}\ref{item:extreme-intersection1}, we know $\ell(C_i)\geq 0$. Consider the two different cycles (say) $C_1,C_2$ that each contain a copy of the edge $(y_t,x_0)$. We know that one of these cycles (say) $C_1$ should have $\ell(C_1)\leq \ell(C)$ and $|C_1|<|C|$. Removing the $(y_t,x_0)$ edge from $C_1$ gives a path (say) $P_1$ such that $(\ell(P), |P|)\succeq (\ell(P_1), |P_1|)$ contradicting the minimality of $P$. 
		
		Now we will show that $Y\Delta Z \in \cI_1 \Rightarrow Y\Delta P \in \cI_1$. First, observe that $r_{\cM_1}(Y\cup Z)=|Y|$ because $r_{\cM_1}(Y+ z)=r_{\cM_1}(Y)=|Y|$ for every $z\in Z$ as we showed $Z\cap X_1=\emptyset$ in \Cref{proof:lem-structural-2a}. On the other hand, the rank $r_{\cM_1}(Y\cup P)\geq |Y|+1$ because $Y\cup P \supseteq Y+x_0 \in \cI_1$. Since $Y \Delta Z \subseteq Y\cup P$, there is an element in $\{x_0,y_1,y_2,\dots,y_{t-1}\}$ that extends $Y\Delta Z$. This has to be $x_0$ because we know $r_{\cM_1}(Y\cup Z)=|Y|$. This implies that 
		\begin{align*}
			(Y\Delta Z) +x_0 = (Y\Delta P) +y_t \in \cI_1\Rightarrow (Y\Delta P)\in \cI_1.
		\end{align*}
		
		\item  ($Y\Delta P \in \cI_2$) Finally, we show that $Y\Delta P\in \cI_2$. Let $N$ be the set of matching edges $(y_i,x_i)$ corresponding to $M_1$ for $1\leq i \leq t-1$ along with $(y_t,x_0)$. Let $M$ be the set of matching edges $(x_{j-1}, y_j)$ corresponding to $\cM_2$ for $1\leq j\leq k$. If $Y\Delta P\notin \cI_2$, then there is an alternate matching $M'$ from vertices in $P\backslash Y$ to $Y\backslash P$. Consider the multiset union of edges $M\cup M'\cup 2N$. Similar to the argument in parts~(1) and~(2a) above, we find a cycle (say) $C_1$ containing the edge $(y_t,x_0)$ such that $\ell(C_1)\leq \ell(C)$ and $|C_1|<|C|$. Deleting the edge $(y_t,x_0)$ gives a path $P_1$ that contradicts the minimality of $P$.  
		\qedhere
	\end{enumerate}
\end{proof}

\Cref{lem:minimal-path-cycle} allows us to compare lengths of minimal
paths using only comparisons of matroid intersections:
\[
\ell(P_1) - \ell(P_2) = w(Y \triangle P_1) - w(Y \triangle P_2),
\]
when $P_1$ and $P_2$ are minimal paths or cycles. In order to optimize over minimal paths, we modify the Bellman-Ford algorithm in the natural way.

\paragraph{Modified Bellman-Ford Algorithm.}
Let $p_y^{(t)}$ denote the minimal path in $\cP(y,t)$. We compute it recursively:
\begin{itemize}
	\item \textbf{Initialization:} For each $y \in Y$,
	\[
	p_y^{(1)} = \min_{x \in X_1} (x \rightarrow y).
	\]
	
	\item \textbf{Update:} For $t \geq 1$,
	\[
	p_y^{(t+1)} = \min \left\{ p_y^{(t)},\; p_z^{(t)} \rightarrow x \rightarrow y \right\},
	\]
	where $(z,x) \in Y \times (E \setminus Y)$ satisfies:
	\begin{enumerate}
		\item $(z,x)$ and $(x,y)$ are arcs in $H$, and
		\item $Y \triangle p_z^{(t)} + x - y \in \cI_1 \cap \cI_2$.
	\end{enumerate}
	\item \textbf{Shortest Augmenting path:} After computing paths $p_y^{(t)}$ for all $y \in Y$ and $t = |Y|$,  we extract a minimal shortest path $P$ from $X_1$ to $X_2$ such that $Y \triangle P \in \cI_1 \cap \cI_2$, and update $Y \gets Y \triangle P$.
\end{itemize}

\begin{proof}[Proof of \Cref{thm:matroid-intersection}]
	Let $Y_t$ be the extreme common independent set of size exactly $t$ starting with $Y_{0}=\emptyset$. Use the modified Bellman-Ford Algorithm to find the shortest augmenting path $P_t$ in the exchange graph $H(\cM_1, \cM_2, Y_t))$. Using \Cref{lem:extreme-intersection} part \ref{item:extreme-intersection1}, the exchange graph has no negative cycles. The shortest path computation takes $O(n^3)$ comparison queries and time (see \cref{sec:shortest-paths}). After every update, we know that $Y_{t+1}:=Y_t \Delta P_t$ is an extreme set using \Cref{lem:extreme-intersection} part \ref{item:extreme-intersection2}. We continue for $n$ steps as long as there is a directed path from $X_1$ to $X_2$ in the exchange graph. Using \Cref{lem:termination}, we know that the maximum cardinality intersection is reached when an augmenting path is not present anymore. Finally, we can compare the weights of all the extreme sets obtained to output the minimum weight common intersection $O(n^4)$ comparisons and $O(n^4)$ time. 
\end{proof}

\subsection{Shortest paths}    
\label{sec:shortest-paths}

Finally, when the combinatorial objects of interest are $s-t$ walks,
we show that the Bellman-Ford shortest path algorithm can be used to
find the shortest $s-t$ walk (which will be a path) using $O(n^3)$
many $s-t$ walk comparisons.

\begin{restatable}[$s$-$t$ walks]{theorem}{stpaths}
	\label{thm:s-t-walks}
	There is an algorithm that finds the minimum-length $s$-$t$ path in
	a graph $G$ (or a negative cycle, if one exists) using $O(n^3)$ 
	$s$-$t$ walk comparisons and $O(n^3)$ time.
\end{restatable}

\begin{proof}
	Assume that for every vertex $v \in V$, there exists a path from $s$ to $v$ and from $v$ to $t$, since otherwise no $s$-$t$ walk can pass through $v$, and we may ignore such vertices. For each $v \in V$, fix an arbitrary $v$-$t$ walk $t_v$. Let $s_v^{(k)}$ denote the shortest $s$-$v$ walk with at most $k$ edges, where $s_v^{(0)} = \emptyset$.
	
	Define the recurrence:
	\begin{align}
		\label{eqn:bellman-ford}
		s_v^{(k+1)} = \min_{u \in V} \left( s_u^{(k)} \circ (u,v) \right),
	\end{align}
	where ties are broken arbitrarily. We can compute this because for any two walks $s_{u_1}^{(k)} \circ (u_1,v)$ and $s_{u_2}^{(k)} \circ (u_2,v)$, we can compare them by extending each with $t_v$ and comparing total weights.
	
	If $w$ has no negative cycles, then $s_t^{(n-1)}$ is the shortest $s$-$t$ path. To detect a negative cycle, check whether $w(s_u^{(n)}) < w(s_u^{(n-1)})$ for any $u \in V$.
	
	Since there are $O(n^2)$ entries $s_u^{(k)}$, and each is computed by comparing $O(n)$ candidate walks, the total number of comparisons is $O(n^3)$. Although each comparison naively takes $O(n)$ time, we can amortize this using shared subwalks, leading to $O(n)$ time per table entry and total runtime $O(n^3)$.
\end{proof}

{\small
  \bibliographystyle{alpha}
  \bibliography{bibdb}
}

\end{document}